# Co-leading Teams Drive Scientific Novelty in Large-scale Research Infrastructures

**Authors**

Mingze ZHANG[1,2,3], Yizhan LI[2,3], Hao PENG[1,*], and Zexia LI[2,3,*]

[1]*Department of Data Science, City University of Hong Kong, Hong Kong SAR, China*
[2]*National Science Library, Chinese Academy of Sciences, Beijing, China*
[3]*Department of Information Resources Management, School of Economics and Management, University of Chinese Academy of Sciences, Beijing, China*

[*]*Corresponding author: Hao PENG* (haopeng@cityu.edu.hk); *Zexia LI* (lizexia@mail.las.ac.cn)

**Abstract**

Large-scale research infrastructures (LSRIs) have become the engine of modern scientific discovery. While these big machines predominantly operate under a user-oriented model where external teams conduct research with support from in-house researchers, the structural integration of staff scientists into user teams and its association with scientific novelty remains unclear. By leveraging a dataset of 273,109 publications across 76 global LSRIs and applying a hybrid machine-learning framework to classify papers into three collaboration patterns: external user only, staff participating, and staff co-leading, we find a distinct novelty premium for external teams that formally integrate staff as co-authors, especially when staff scientists play co-leading rather than participating roles. Further, the premium peaks at a relatively balanced user-staff team composition, potentially due to an "epistemic lock-in" by either party. Crucially, we find that the ideal collaboration architecture evolves with user experience: while newcomers can obtain a large novelty premium from mere staff participation, experienced users only benefit from staff co-leading teams. This result suggests a "knowledge saturation effect" for which a deeper intellectual partnership is needed to sustain novelty. By revealing how user-staff collaboration structure drives scientific creativity, our study offers practical policy implications for the strategic management and intervention of LSRIs in the era of human-machine collaboration.

## 1 Introduction

The landscape of modern scientific discovery is increasingly driven by Big Science (Bianco et al., 2017; Börner et al., 2021), characterized by huge, capital-intensive, and highly complex experimental facilities (Hallonsten, 2016a; Heinze & Hallonsten, 2017) known as large-scale research infrastructures (LSRIs). From the Large Hadron Collider at CERN (Brumfiel, 2011) to advanced synchrotron light sources (Conroy, 2024; Jiménez, 2010), these infrastructures push the absolute boundaries of human knowledge by focusing on cutting-edge research projects (Zhang, Li, et al., 2026; Zhang, Wang, et al., 2025). Crucially, the governance of most contemporary LSRIs has shifted toward a user-oriented model since the end of Cold War (Hallonsten & Christensson, 2017). Under this model, the core scientific output is not generated by massive, in-house teams, but rather by small groups of external academic users who compete for limited experimental beamtime to test their frontier hypotheses (D'Ippolito & Rüling, 2019; Silva et al., 2019).

Within this user-oriented model, the micro-dynamics of knowledge production heavily depend on the interaction between two distinct groups (Anupama et al., 2021; Söderström, 2023b): external principal investigators (who bring domain-specific scientific inquiries) and in-house scientists (who possess deeply embedded tacit knowledge of the facility's cutting-edge capabilities). While the broader literature on facilitymetrics has documented the motivations for these collaborations and how they influence the downstream scientific outcomes, the structural division of labor during the collaborations remains a critical blind spot. One foundational qualitative study, notably (D'Ippolito & Rüling, 2019), has provided invaluable ethnographic insights into these collaborations, revealing that in-house scientists can act as anything from marginal technical service providers to genuine peer collaborators. However, the literature critically lacks a large-scale empirical validation of how these nuanced collaboration archetypes are associated with the epistemological outcomes of the research.

This study bridges this gap by addressing a fundamental question: How does the structure of in-house scientist involvement in external user teams affect the production of novel scientific knowledge? Methodologically, we introduce a scalable computational approach to the emerging field of facilitymetrics. Building on the hybrid machine-learning framework pioneered by F. L. Xu et al. (2022), we adopt and fine-tune a Dense Neural Network (DNN) classifier alongside Large Language Models (LLMs) to categorize the division of scientific labor (Leadership vs. Support roles) across hundreds of thousands of LSRIs publications (Y. Lin et al., 2023; Seo & Bu, 2025). Further, by combining these identified team roles with author roles (external users vs. in-house scientists) derived from affiliations (Silva et al., 2019; Zhang, Wang, et al., 2025), this methodological pipeline enables us to operationalize three collaboration patterns and quantitatively test their novelty premium at a global scale across approximately 270,000 publications from 76 global LSRIs.

Our econometric analyses yield three findings. First, we identify a measurable novelty premium for formal user-staff collaboration: external user teams that integrate in-house scientists as co-authors produce statistically higher scientific novelty, particularly when external users share leadership with in-house scientists. Second, we uncover a robust inverted U-shaped relationship between the proportion of in-house scientists and scientific novelty, both among the whole group members and within its leadership team. This provides suggestive evidence for the risk of "epistemic lock-in" where unbalanced team compositions can direct the research toward conventional routines. Finally, we find an evolutionary shift in collaborative needs as external users accumulate facility experience and staff ties: while novice users receive a larger novelty premium from the foundational support of mere staff participation, this benefit depreciates for experienced users, who instead require deeper intellectual partnerships via staff co-leadership. Collectively, our study advances the facilitymetrics and team science literature by demonstrating that the optimal collaboration pattern often emerges from a balanced user-staff team composition and that the ideal architectural design required to produce novel contribution within LSRIs is not static but dynamically evolves with user experience and network ties.

The remainder of this paper is structured as follows. Section 2 reviews the extant literature on Big Science and its large-scale infrastructures, team science, and user-staff collaborations to establish our theoretical foundation. Building on this background, Section 3 develops our research hypotheses concerning the collaboration premium, the risks of cognitive lock-in, and the dynamics of collaboration patterns. Section 4 briefly introduces the data curation pipeline from global LSRIs and our machine-learning classification framework and econometric specifications (Further details are provided in Supplementary Materials). Section 5 presents the empirical results. Section 6 discusses the theoretical contributions and policy implications of our findings, alongside the study's limitations and avenues for future research. Finally, Section 7 provides the concluding remarks.

## 2 Literature Review

### 2.1 Big Science era and scientific tools

As Price (1963) demonstrated, scientific research transitioned into the era of Big Science in the aftermath of World War II. Since then, the demand for advanced analytical technologies has grown exponentially (Heinze & Hallonsten, 2017; Kozheurov & Teimurov, 2020). Today, Big Science constitutes a defining feature of modern research (Crease & Westfall, 2016), characterized by numerous national and supranational projects worldwide (Börner et al., 2021) that rely on huge, highly complex, and capital-intensive facilities (Hallonsten, 2014; Heidler & Hallonsten, 2015). These facilities, known as Large-scale Research Infrastructures (LSRIs), are reported to have become the engine of scientific discovery, particularly in the STEM-related domains (Börner et al., 2021; Manganote et al., 2016). Currently, it is reported that Nobel Prizes show close associations with these infrastructures (Stix, 2001). For instance, the experimental confirmation of the Higgs boson relied heavily on the Large Hadron Collider at CERN (Brumfiel, 2011), and the first direct observation of gravitational waves was made possible by the highly complex LIGO observatories (Castelvecchi, 2017). Similarly, in the life sciences, the high-resolution structural determination of G-protein-coupled receptors—which earned the 2012 Nobel Prize in Chemistry—fundamentally depended on the finely focused X-ray microbeams provided by synchrotron facilities such as the Advanced Photon Source and the European Synchrotron Radiation Facility (Cherezov et al., 2007; Rasmussen et al., 2011). There is no doubt that these huge machines strongly push the limits of human knowledge, placing academia on an endless journey to chase the frontiers of science (Hallonsten & Christensson, 2017).

Currently, most LSRIs around the globe follow a user-oriented model (D'Ippolito & Rüling, 2019; Langford & Langford, 2000). Consequently, Big Science has developed a new feature: research at these facilities is typically conducted by small groups of external users rather than massive, centralized in-house teams (Nogrady, 2023). This paradigm is defined as the “big machine and small team” model by Hallonsten (2016a). These external users—a cohort of scientists predominantly affiliated with academia and rarely from industry—submit research proposals to the facilities they intend to use, awaiting approval and the assignment of experimental beamtime (Yang et al., 2024). Then they conduct on-site experiments at the facility and publish manuscripts based on their observational results. In this context, LSRIs serve as massive experimental infrastructures equipped with cutting-edge technologies, offering advanced capabilities the discovery of novel knowledge (Hallonsten & Christensson, 2017; Heinze & Hallonsten, 2017). However, they represent an exceedingly rare resource for scientific research, as their annual operational capacities are strictly limited (Bianco et al., 2017; Hallonsten, 2016b), which triggers fierce intellectual competition among all potential users (Söderström, 2023a; Zhang, Wang, et al., 2026). To mitigate risks, the assignment of these limited resources tends to prioritize external users with strong academic backgrounds and high reputations, thereby providing assurances of beneficial scientific returns to the facilities (Kozheurov & Teimurov, 2020; Lauto & Valentin, 2013). Furthermore, a recent study verified the positive contributions of these facilities to producing novel knowledge by comparing their outputs with papers produced without such assistance (Zhang, Li, et al., 2026).

### 2.2 Scientific collaboration and team science

Beyond the sheer scale of infrastructure, modern science is distinctly characterized by a pervasive shift toward collaboration (Wuchty et al., 2007). As scientific knowledge becomes increasingly complex, individual researchers are rarely equipped to tackle multifaceted problems alone (Guimerà et al., 2005). Consequently, team has emerged as the fundamental unit of knowledge production across the scientific ecosystem, prompting a wealth of Science of Science research over the past two decades (Jones et al., 2008; Y. Lin et al., 2023; Liu et al., 2022; Wu et al., 2019; F. L. Xu et al., 2022; Yang et al., 2022).

Under the context of Team science and combined with the large-scale bibliographical data analysis, the formation and outcomes of scientific collaborations are widely discussed. Conventionally, co-authors on a single paper are considered a team, and diverse team compositions based on co-authors' inherent backgrounds—such as race (Hofstra et al., 2020; Li et al., 2025), ethnic (Hoogendoorn et al., 2012; Paletz et al., 2004), gender (Yang et al., 2022; M.-Z. Zhang et al., 2024), geography (Hoekman & Rake, 2024; van der Wouden & Youn, 2023), and career stages (H. M. Xu et al., 2022; Xu et al., 2024)—can influence team outputs, which are mainly captured by novelty, disruption, and scientific impact (Thelwall et al., 2024; Wu et al., 2024; Zhang, Wang, et al., 2025). The variations in performance are not only estimated by empirical data but also supported by long-established theories in social sciences. According to the viewpoints of Social Exchange Theory (SET), individuals in one system tend to interact with each other and exchange information, support, and resources to achieve their goals while this theory applies appropriately to academia, unveiling the mechanisms of why scientists collaborate with their peers (Blau, 2017; Cropanzano & Mitchell, 2005). From the perspective of Cognitive Diversity, it is expected that scientific collaboration could yield better performance in knowledge production since heterogeneous resources are combined, communicated, and integrated to spark new ideas (Mansoor et al., 2013; Mello & Rentsch, 2015). However, excessive diversity might be harmful to scientific teams due to sharp increases in coordinative costs and inevitable disagreements on research projects, and such negative effects can be more pronounced in interdisciplinary teams (Cheng et al., 2025; Wang et al., 2022). According to the Similarity-Attraction Theory, individuals could be driven to better performance when they are self-identified in a homogeneous group (Montoya & Horton, 2013). However, for those teams without any boundary spanner, identical knowledge backgrounds and similar network resources could also damage research outcomes (Wells & Aicher, 2013).

This line of literature suggests an inverted U-shaped relationship between team composition and collaborative performance (Liu et al., 2022; Wang et al., 2022; Wu et al., 2024). With scientific advancement increasingly driven by large-scale machines, it is critical to examine the empirical pattern by quantifying the direction and effect size of team composition on scientific novelty.

### 2.3 Collaboration in the context of large-scale research infrastructures

Existing studies in facilitymetrics mainly focus on the interaction between two distinct groups of scientists: external users and in-house scientists (Silva et al., 2019; Söderström, 2023b). Given that these huge machines are predominantly driven by external users to produce new knowledge, the emergence and development of collaborations depend largely on the cognitive frameworks and initiatives of these external scientists (Hallonsten & Christensson, 2017; Heinze & Hallonsten, 2017). In this dynamic, in-house scientist often occupy a relatively passive position (D'Ippolito & Rüling, 2019; Söderström, 2023b). As a result, collaborations within these facilities tend to fall into two primary patterns: those with the involvement of in-house scientists (the internal group) and those without, known as the external group (Zhang, Wang, et al., 2025).

Identifying these two groups is fundamentally a post-hoc process, as it relies on bibliometric co-authorship data to retroactively infer collaboration dynamics (Silva et al., 2019). However, inferring the collaborative process exclusively from authorship records is inherently limited, because, even formal collaborations do not directly equate co-authorship, let alone informal ones (Danús et al., 2026). The observed collaborative outputs of the internal group largely materialize because external teams adhere to academic authorship norms, formally recognizing instances where in-house scientists have made substantial intellectual or technical contributions (D'Ippolito & Rüling, 2019). Beyond the inherent reliance on external users to grant formal recognition, internal group publications fail to constitute the mainstream due to a confluence of other factors: the overarching agency of external users, the constrained bandwidth and energy of these facility staff, as well as the broader limitations in facility-level policies and resources (Hallonsten & Christensson, 2017; Lauto & Valentin, 2013).

While prior research has explored user-staff team dynamics within specific facilities (M. ZHANG et al., 2024; Zhang, Wang, et al., 2025), these investigations are fundamentally constrained by their small-scale, localized scope and surface-level treatment of collaboration. Crucially, the literature lacks a large-scale, global empirical analysis to systematically evaluate these collaboration patterns across diverse infrastructures. Furthermore, a nuanced empirical understanding of *how* in-house scientists should engage (e.g., merely as supporters vs. project co-leaders) and *when* their intervention is most effective (e.g., during a user's initial exposure to the facility vs. after the accumulation of sufficient experiential knowledge) remains critically absent.

## 3 Hypotheses

### 3.1 Novelty Premium for formal collaboration: In-house scientists as boundary spanners

In the context of large-scale research infrastructures, scientific knowledge production relies heavily on the integration of heterogeneous resources. Building on the concepts of cognitive diversity (Mello & Rentsch, 2015) and Social Exchange Theory (Blau, 2017), diverse teams are better equipped to recombine distant knowledge domains. External users typically bring domain-specific theoretical expertise and research questions, whereas in-house scientists possess deep, tacit knowledge regarding the facility's cutting-edge capabilities and methodological boundaries (D'Ippolito & Rüling, 2019). When in-house scientists are formally integrated into external user teams, they act as critical boundary spanners, bridging the gap between scientific inquiries and machine capabilities (Zhang, Wang, et al., 2025). Therefore, moving beyond pure external user teams by involving in-house scientists in teams can provide a distinct novelty premium.

***Hypothesis 1 (H1):*** *Compared to teams purely composed of external users that only receive marginal support from in-house scientists, teams that formally integrate in-house scientists as co-authors exhibit a higher probability of producing novel scientific knowledge.*

### 3.2 Epistemic Authority: The leadership role of in-house scientists

The depth of in-house scientists' involvement might significantly alter the novel knowledge outputs. Three distinct collaboration patterns, pure external user teams without formal authorships to in-house scientists, teams featuring the mere participation of in-house scientists, and teams characterized by joint leadership, could be justified in the context of LSRIs. When in-house scientists share team leadership rather than acting strictly as supporting co-authors, they gain epistemic authority to reshape the research trajectory (Xu et al., 2024). This shared leadership allows for a more symmetrical exchange of ideas, where the technological potential of the facility can be fully leveraged to address the external users' frontier questions, thereby maximizing the novel potential of the research (Lauto & Valentin, 2013).

***Hypothesis 2 (H2):*** *Teams co-led by external users and in-house scientists are associated with higher novelty than teams with in-house scientists merely participating in a supporting role.*

### 3.3 Epistemic Lock-in: The risks of imbalanced team composition

While in-house scientists provide essential technical and cognitive benefits, the structural composition of the team must be carefully calibrated to maximize scientific novelty. The proportion of in-house scientists, both within the overall team and its leadership, might exhibit an inverted U-shaped relationship with novelty (Wu et al., 2024), driven by the dual risks from imbalance collaboration structure and its downstream "epistemic lock-in".

On one end of the spectrum, if external users disproportionately dominate the collaboration with minimal in-house integration, the team may suffer from insufficient structural and epistemic support from facility staff, making it difficult to fully unlock the infrastructure's cutting-edge capabilities for novel discoveries. On the opposite end, excessive reliance on facility staff might trigger equally detrimental consequences. According to the Similarity-Attraction Theory (Montoya & Horton, 2013), overwhelming staff homogeneity may stifle users' unconventional idea, prioritizing to research routines and conventional methodologies at the expense of risky, atypical knowledge recombination. Therefore, optimizing novelty might require maintaining a delicate equilibrium where neither party completely subordinates the other.

***Hypothesis 3a (H3a)****: The proportion of in-house scientists within a team exhibits an inverted U-shaped relationship with the production of novel knowledge, underscoring the necessity of maintaining a structural balance between external users and facility staff, as both the under-representation of in-house expertise and the over-embeddedness of institutional routines hinder collaborative innovation.*

***Hypothesis 3b (H3b)****: The proportion of in-house scientists sharing team leadership exhibits an inverted U-shaped relationship with the production of novel knowledge, highlighting the critical need for maintaining an epistemic balance in user-oriented facilities.*

### 3.4 New vs. Repeated Ties: Relational dynamics and diminishing returns

The value of user-staff collaboration is not static but evolves alongside the external users' familiarity with these machines. As external users accumulate experiential knowledge at a specific facility, their reliance on the boundary-spanning capabilities of in-house scientists might decrease. The external users' prior experience at the facility and prior knowledge exchanges with staff can act as crucial moderating variables, attenuating the performance advantages brought by the involvement of in-house scientists. For external users engaging with a facility for the first time, in-house scientists are indispensable for navigating the facility's complexities (D'Ippolito & Rüling, 2019). However, as external users accumulate experiential knowledge and repeatedly interact with the facility and its staff, they gradually internalize these cognitive and technical capabilities. Therefore, for highly experienced users, repeated ties with in-house scientists may associate with cognitive fatigue rather than complementary insights (Liu et al., 2024; Liu et al., 2022; Santos et al., 2024). Prolonged over-reliance on these staff might trap the team in established research routines, thereby diminishing the novelty premium.

***Hypothesis 4a (H4a):*** *External users' accumulated publication volume at the focal facility negatively moderates the relationship between in-house scientist involvement and scientific novelty.*

***Hypothesis 4b (H4b):*** *External users' prior collaborative ties with in-house scientists negatively moderate the relationship between in-house scientist involvement and scientific novelty.*

## 4 Data and study design

### 4.1 Data curation

To test the proposed hypotheses, compiling a robust dataset of publications supported by large-scale research infrastructures (LSRIs) was the critical first step. Following the established framework of Facilitymetrics (Silva et al., 2019; Söderström et al., 2022; Zhang, Wang, et al., 2025), we collected the BSFs' publications from self-constructed databases in those facilities' official websites rather than retrieving them from commonly used bibliographical databases such as Web of Science or Scopus (Please refer to Supplementary Text 1 Dataset collecting procedure for details). Since most LSRIs are publicly funded infrastructures (Hallonsten, 2014), they are typically mandated to showcase their scientific achievements to taxpayers and funding agencies (Börner et al., 2021; Hallonsten, 2013). Consequently, these facilities systematically collate and publish their research outputs on their official websites (Giffoni & Florio, 2023).

Retrieving LSRI-supported publications exclusively from standard bibliographic databases risks severe data loss and selection bias. In such databases, facility identification relies heavily on acknowledgment texts, which are notoriously unstandardized and frequently omitted (D'Ippolito & Rüling, 2019) or affiliation addresses. Querying strictly by affiliation captures only those publications involving in-house scientists, thereby systematically overlooking the vast majority of outputs produced purely by external users (Silva et al., 2019; Söderström, 2023a; Zhang, Wang, et al., 2026).

Between April and May 2025, we curated a master list of global LSRIs, guided by prior research and expert recommendations from the Chinese Academy of Sciences. We successfully collected publication data from 88 mainstream facilities, selected based on their research tiers and the accessibility of their publication repositories (See Supplementary Table S11). Given that metadata structures vary drastically across different facility websites, it was essential for us to introduce large-scale bibliographical database as supplementary data to overcome the limitations of websites' data, We therefore applied OpenAlex to finish our research design (Priem et al., 2022), which is known as a fully open-access database for Science of Science research that has been used widely across the global community (Peng et al., 2024; Xing et al., 2025; Xu et al., 2024). Our version of OpenAlex was updated in Jan. 2025.

After matching with OpenAlex, we finally retained 315,951 published records involving 76 user-oriented facilities but there were only unique 273,109 publications, with the discrepancy indicating widespread facility co-utilization (Zhang, Wang, et al., 2026). More details are provided in Supplementary Text 1 Dataset collecting procedure.

### 4.2 Measurements

#### 4.2.1 Scientific novelty as the measurement of team performance

We applied atypical combinations of journals to quantify the scientific novelty, which proposed by Uzzi et al. (2013) and has been reproduced by numerous previous studies in the domain of Science of Science (Ke et al., 2026; Teplitskiy et al., 2022; Wagner et al., 2019; Yang et al., 2022). This metric, rooted in Schumpeterian innovation theory (Malerba & Orsenigo, 1995), posits that atypical combinations of prior knowledge mean novelty while common combinations demonstrate the conventionality of scientific papers, and the combinations are represented by co-cited journals. The differences, captured by the z-score, between the observed frequency and the expected frequency for every journal combination, normalized by the standard errors according to the simulated co-citation network, demonstrate whether the combination is uncommon or not. For every publication with a few journal references, a z-score distribution can be accessed, and 10$^{th}$ z-score is considered to represent the novelty score (NS) of the publication. Following the previous work, a novel publication is defined as a publication with NS lower than zero and we therefore adopted this method and accessed the original novelty scores of every publication in our dataset from SciSciNet-v2 (Z. Lin et al., 2023).

#### 4.2.2 Quantifying team collaboration patterns

To systematically uncover the division of scientific labor between in-house scientists and external users in the context of LSRIs, we introduced a scalable computational approach to the field of facilitymetrics. Building on the machine-learning framework pioneered by F. L. Xu et al. (2022), we adopted and finely tuned a Dense Neural Network (DNN) classifier to categorize authors' roles in teams, subsequently identifying in-house scientists based on affiliation data (Zhang, Wang, et al., 2025). Detailed technical specifications, data processing steps, and model evaluations are provided in Supplementary Text 2 Collaboration pattern identifications.

In summary, we utilized a dataset of 95,014 author contribution statements from F. L. Xu et al. (2022) to construct our ground truth data. After extracting the natural language verbs describing each author's contribution, we mapped these records into the OpenAlex database, yielding over 630,000 valid paper-author pairs. To categorize these contributions at scale (Seo & Bu, 2025), we employed a Large Language Model (DeepSeek-R1) using a few-shot prompting strategy (Prompt are recorded in Supplementary Text 2.2.1 Access to response variable). Authors' tasks were conceptually aggregated into a binary framework: Leadership (e.g., conceptualizing, directing, writing (Lu et al., 2020)) and Support (combining direct operational tasks and indirect advisory roles (Y. Lin et al., 2023; Lu et al., 2022)). Evaluated against a manually annotated benchmark, the LLM demonstrated high accuracy, achieving an F1-score of 0.97 for identifying leadership roles.

To extend this classification to our full dataset of publications supported by large-scale research infrastructures, we trained a Dense Neural Network classifier. We calculated 13 bibliometric predictor variables (Detailed in Supplementary Text 2.2.2 Access to predictor variables) using OpenAlex data to capture an author's contribution to both the focal paper (e.g., author order (Sauermann & Haeussler, 2017; Tscharntke et al., 2007), fractional contribution (M.-Z. Zhang et al., 2024), reference contributions) and the broader scientific system (e.g., career age (Duan et al., 2025), past topological diversity (Tian et al., 2025; Zeng et al., 2022), citation impact (Yoo et al., 2024)). The optimized DNN achieved an F1-score of 0.77, providing a robust and scalable mechanism to predict whether an author served in a leadership or support capacity in any given scientific team (See details in Supplementary Text 2.2.3 Model training).

The final step involved linking the predicted team roles with authors' affiliations (Söderström, 2023b). We identified in-house scientists by matching authors' affiliation records in OpenAlex with the unique Research Organization Registry (ROR) IDs of the respective large-scale facilities or their parent national laboratories (Zhang, Wang, et al., 2026) (Details are provided in Supplementary Text 2.3 The classification of in-house scientists).

By intersecting the authors' institutional affiliations (in-house vs. external) with their predicted team roles (leadership vs. support), we systematically mapped the team structure into one informal and two formal collaboration patterns: *Only external users teams (without in-house scientists co-authors)*, *teams with In-house scientists participating*, and *teams with In-house scientists co-leading* (D'Ippolito & Rüling, 2019). This categorization moves beyond simple co-authorship counts, allowing us to quantitatively evaluate how the nature and depth of in-house scientist involvement drives knowledge production (See details in Supplementary Text 2.4 The identification of collaboration pattern). Notably, under the user-oriented model, publications from LSRIs are mainly contributed by teams purely composed of external users (See the annual distributions in Supplementary Figure S7).

4.2.2 Control variables

To isolate the net effect of collaboration pattern on novel knowledge production, we control for a comprehensive set of confounding factors across three dimensions: paper-level features, team-level factors, and external leader characteristics (All details are provided in Supplementary Text 3 Constructing confounding variables in regressions).

*Paper-level controls*: We account for the research scale and scope by controlling for the number of references (log-transformed) (Uzzi et al., 2013) and whether the publication is an SDG paper (a dummy indicating relevance to Sustainable Development Goals, utilizing OpenAlex's Aurora-based classifier), as sustainability research associates with higher novelty in the context of large-scale research infrastructures. We also control for Facility co-utilization (Zhang, Wang, et al., 2026), counting the number of distinct facilities used for the paper, to account for the complexity of the technical infrastructure.

*Team-level controls*: We control for Team size (log of unique authors) (F. L. Xu et al., 2022; Yang et al., 2022) and International collaboration (a dummy for multi-country authorship) (Ke et al., 2026). To account for the team's overarching seniority, we include Average team age (Xu et al., 2024), calculated as the log-transformed average career age (years since first publication) of all co-authors.

*External leader characteristics*: We control for the academic capital and geographic proximity of the external leadership (Ke et al., 2026; Liu et al., 2022; H. M. Xu et al., 2022). This includes their average historical citation impact (log-transformed) and the prestige of their affiliated institutions (measured by the average institutional h-index, log-transformed). We include a Global North dummy (1 if $\geq$ 50% of external leaders are from Global North countries[1]) and a Same Country dummy (1 if $\geq$ 50% are domestic to the facility) to control for geographic and economic advantages (Zhang, Lyu, et al., 2025). Finally, we account for cognitive proximity via External leader knowledge similarity, a continuous variable measuring the cosine similarity between the semantic vectors (generated via the SPECTER 2.0 model (Cohan et al., 2020)) of the leaders' prior publications and the facility's prior outputs in the same field (Hao et al., 2026; Yang et al., 2026). Additionally, we introduce two moderating variables, which are detailed as follows.

4.2.4 Moderating variables

To understand how the effectiveness of different collaboration patterns varies by the experience of external leaders (D'Ippolito & Rüling, 2019; Lauto & Valentin, 2013), we introduce two moderating variables that capture external leaders' past experience with the focal facility and their embeddedness with its staff scientists.

*Prior facility experience*: Calculated as the average number of prior publications the team's external leaders have produced using the focal facility. It proxies the external leaders accumulated technical familiarity with the infrastructure.

*Prior in-house collaboration*: A dummy variable equal to 1 if any external leader in the focal team has a historical co-authorship record with the facility's in-house scientists prior to the focal publication, and 0 otherwise. This captures pre-existing knowledge exchanges between external users and in-house scientists.

## 4.3 Econometric estimation

[1] The classification of Global South is determined by the list provided by: https://www.g77.org/, and countries/regions out of the list are classified as Global North.

To estimate the impact of user-staff collaboration patterns on the production of novel knowledge, we employ a two-way fixed-effects regression model. The baseline specification is formally defined as:

$$(1)\quad Novel\ paper_{i,d,t} = \alpha + \beta_1 CoType_{i,d,t} + \gamma X_{i,d,t} + \mu_d + \tau_t + \varepsilon_{i,d,t}$$

Where $Novel\ paper_{i,d,t}$ represents the knowledge novelty of focal paper $i$ in research discipline $d$ and published in year $t$. $CoType_{i,d,t}$ is the categorical independent variable indicating the identified collaboration structures (informal type: *only external user teams (without in-house scientists co-authors)*, formal types: *teams with In-house scientists participating*, and *teams with In-house scientists co-leading*), with the informal type serving as the reference group. The vector $X_{i,d,t}$ represents the suite of confounding variables detailed in Sections 4.2.3 and 4.2.4. To rigorously account for unobservable heterogeneity across scientific domains and temporal trends, we incorporate two-way fixed effects. Specifically, $\mu_d$ denotes discipline fixed effects based on the 26 research fields from OpenAlex (allowing up to three disciplines per paper), absorbing field-specific citation conventions and research paradigms under the context of large-scale research infrastructures. $\tau_t$ denotes publication year fixed effects to control for time-variant macroeconomic shocks and shifts in global scientific funding. $\varepsilon_{i,d,t}$ is the robust error term.

In subsequent models, we introduce interaction terms between $CoType_{i,d,t}$ and our moderating variables to test for moderating effects. Additionally, we firstly use a binary independent variable (Without vs. With in-house scientists, merged by the last two patterns) for H1.

## 5 Results

### 5.1 Co-leading teams drive scientific novelty among formal user-staff collaboration.

To empirically test H1 and H2, we first examine the net effects of in-house scientist involvement on the likelihood of producing novel scientific outputs. Notably, the independent variable is a binary indicator capturing the presence of any in-house scientist for H1. For H2, we utilize the categorical variable (*CoType*) to distinguish specific collaboration patterns. In both models, teams consisting purely of external users serve as the reference group.

As reported in Table 1 Column (1), the binary indicator for the involvement of in-house scientists in a team is positive and statistically significant ($\beta = 0.065$, $p < 0.01$). The marginal effects illustrated in Figure 1a visually corroborate this finding (see the orange node and error bars), demonstrating that external user teams formally integrating in-house scientists possess a notably higher predicted probability of generating novel knowledge compared to purely external user teams (average marginal probabilities: 37.2% vs. 35.8%, representing a 1.04-fold increase, 95%CI = [1.03, 1.05]). This provides strong empirical support for H1, confirming that in-house scientists involved as co-authors is associated with a distinct novelty premium by acting as boundary spanners between scientific inquiries and complex facility capabilities.

Column (2) of Table 1 disaggregates this formal involvement into two distinct collaboration patterns: the mere participation of in-house scientists and shared leadership. The results reveal a stark contrast: while the coefficient for in-house scientists in a co-leading role is highly significant and positive ($\beta = 0.094$, $p < 0.01$), the coefficient for mere participation is statistically insignificant ($\beta = 0.002$, $p = 0.901$). The pink and red nodes and error bars in Figure 1a further highlight this divergence, showing that the novelty premium is primarily driven by teams where external users share leadership with in-house scientists (predicted novelty probability is 37.9%, representing a 1.06-fold increase, 95%CI = [1.04, 1.07]). This supports H2, indicating that informal collaboration that without co-authorship and staff serve as supporting roles in teams (predicted novelty probability is 35.9%, representing a 1.00-fold increase, 95%CI = [0.98, 1.02]) are insufficient for novel knowledge; rather, in-house scientists must be granted epistemic authority to truly leverage the facility's potential for novel breakthroughs.

We further conduct Propensity Score Matching (PSM) to control for the selection bias and compare the Average Treatment Effect on the Treated (ATT) with 95% confidence intervals to validate H1 and H2. Noted that we take teams consisting purely of external users as the control group and conduct matching three times, comparing them sequentially against teams with any in-house scientists as co-author, teams with in-house scientists in a merely participatory role, and teams sharing leadership with in-house scientists (Details of balance checks can be found in Supplementary Table S5 to Supplementary Table S7, where most differences in covariates between two groups are insignificant). The visualized results of PSM can be seen in Figure 1b. According to the bootstrapped samples, the ATT is insignificant for teams with in-house scientist mere participation (0.6%, 95%CI = [-0.2%, 1.4%]) while teams formal with in-house scientists (2.0%, 95%CI = [1.5%, 2.4%]) and teams sharing leadership with in-house scientists (1.8%, 95%CI = [1.3%, 2.4%]) show positive relationships with the probability of producing novel knowledge.

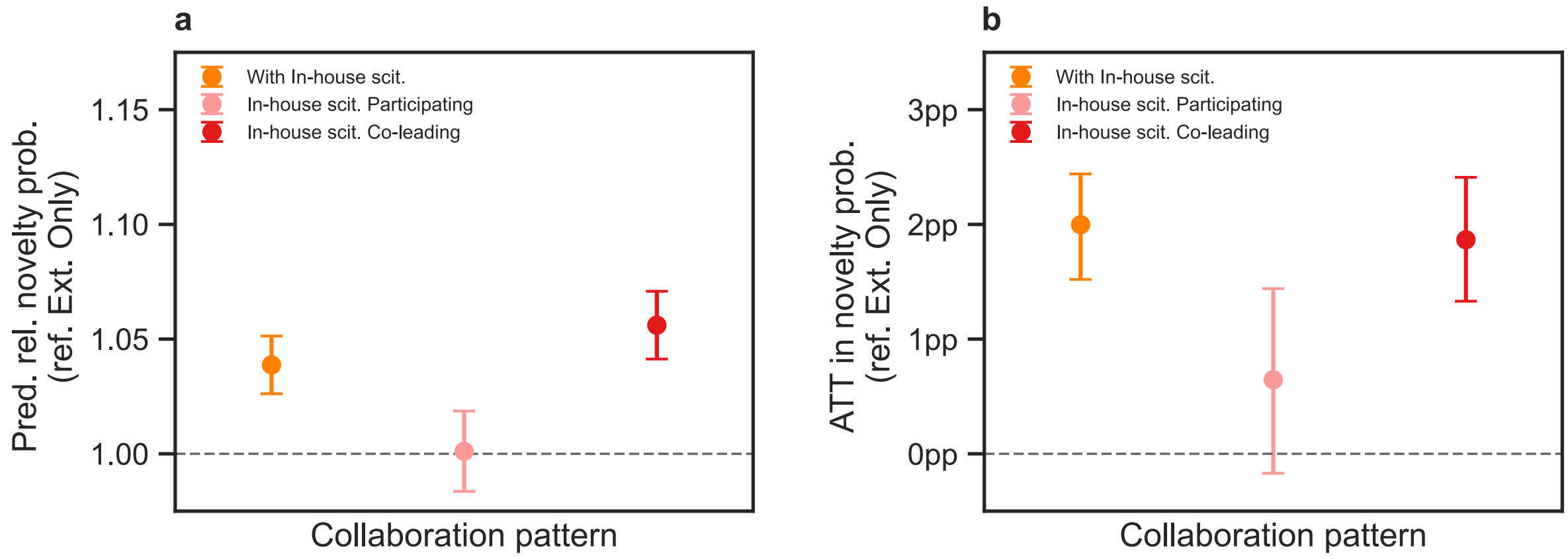

**Figure 1. The formal involvement of in-house scientists in external user teams is associated with a higher probability of producing novel research and the novelty premium is mainly driven by co-leading teams. a,** Predicted relative probabilities of scientific novelty for external user teams with any in-house scientists as co-author (orange bar), which can be further broken down into (i) teams with in-house scientists in participatory roles (pink bar) and (ii) teams co-led by in-house scientists (red bar). Marginal effects are predicted based on regression models shown in Table 1 column (1) and column (2). **b,** The net effect of collaboration patterns on team novelty estimated based on the Propensity Score Matching (PSM). The Average Treatment Effect on the Treated (ATT) robustly supports the observed novelty gaps between teams with and without in-house scientists (reference group). Balance checks are provided in Supplementary Table S5 to Supplementary Table S7. Error bars represent 95% standardized (in **a**) or bootstrapped confidence intervals (in **b**).

**Table 1. Logistic regression models for predicting the likelihood of producing novel scientific knowledge based on in-house scientist involvement and controls.** Both models use teams consisting purely of external users as the baseline reference group. All specifications control for paper characteristics, external leader attributes, prior facility experience, and fixed effects for research discipline and publication year. Column (1) employs a binary independent variable indicating the presence of any in-house scientist as co-author. Based on our role classification framework, Column (2) disaggregates this formal involvement into mere participation and shared co-leadership. Step-wised regressions results are provided in Supplementary Tables S3 and S4. Robust errors are reported in parentheses. Significance levels are based on two-sided Wald tests: *p<0.1, **p<0.05, ***p<0.01.

| | **DV: Novel paper = True** | |
|---|---|---|
| | (1) | (2) |
| **With in-house scit. = True** | 0.065*** | |
| | (0.010) | |
| **In-house scit. participating = True** | | 0.002 |
| | | (0.015) |
| **In-house scit. co-leading = True** | | 0.094*** |
| | | (0.011) |
| **Num. author (log)** | -0.088*** | -0.083*** |
| | (0.007) | (0.007) |
| **International paper = True** | -0.046*** | -0.047*** |
| | (0.010) | (0.010) |
| **Num. references (log)** | 0.053*** | 0.053*** |
| | (0.009) | (0.009) |
| **Num. facilities** | 0.067*** | 0.066*** |
| | (0.007) | (0.007) |
| **SDG paper = True** | 0.084*** | 0.084*** |
| | (0.008) | (0.008) |
| **Avg. team age (log)** | 0.011 | 0.009 |
| | (0.013) | (0.013) |
| **Avg. ext. leader avg. impact (log)** | -0.275*** | -0.275*** |
| | (0.007) | (0.007) |
| **Avg. ext. leader inst. h-index (log)** | -0.050*** | -0.049*** |
| | (0.008) | (0.008) |
| **Ext. leader Global North = True** | 0.309*** | 0.309*** |
| | (0.018) | (0.018) |
| **Ext. leader Same Country = True** | -0.060*** | -0.061*** |
| | (0.009) | (0.009) |
| **Ext. leader knowledge similarity** | -0.802*** | -0.806*** |
| | (0.160) | (0.160) |
| **Ext. leader avg. prior pub in Facility (log)** | -0.042*** | -0.041*** |
| | (0.005) | (0.005) |
| **Ext. leader prior with in-house scit. = True** | 0.331*** | 0.330*** |
| | (0.011) | (0.011) |
| ***Fixed-effects controls*** | | |
| **Disciplines (26 fields)** | Yes | Yes |
| **Publication year (51 years)** | Yes | Yes |
| **Obs.** | 294,879 | 294,879 |
| **Pseudo $R^2$** | 0.059 | 0.060 |

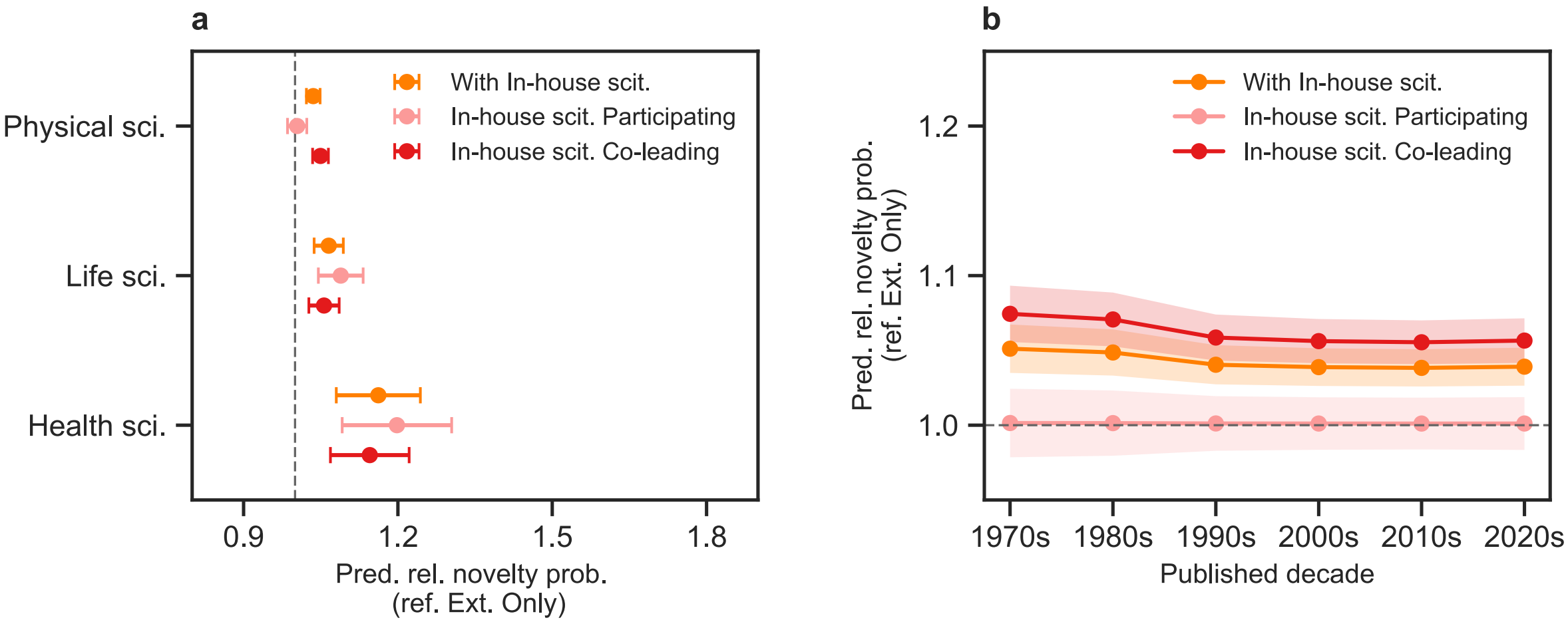

**Figure 2. Consistent novelty premium of formal user-staff collaboration across research domains and time.** **a,** Predicted relative probabilities of scientific novelty for distinct types of formal user-staff collaboration (teams consisting purely of external users are treated as the reference group) across three different research domains. Given that most large-scale research infrastructures primarily support the Physical Sciences rather than the Social Sciences (See Supplementary Table S1 for publication volumes), we therefore exclude the studies in Social Sciences. See Supplementary Table S8 and Supplementary Table S9 for the corresponding regression results. **b,** Predicted relative probabilities of scientific novelty for three nested types of formal user-staff collaboration (teams consisting purely of external users are treated as the reference group) across six decades. Error bars represent 95% standardized confidence intervals.

These findings remain largely consistent across different research domains (Figure 2a for visualized results and Supplementary Table S8 (binary IV) - Supplementary Table S9 (*CoType* IV) for regression results). Notably, according to the publication volumes in Supplementary Table S1, research related to the Physical sciences is dominant, while studies originated from Life sciences and Health sciences are not the mainstream in the context of LSRIs. Moreover, studies related to Social Sciences are further rare, so we excluded them in this analysis. However, the results show that the coefficients and effect sizes of in-house scientist formal involvement on novel probability are relatively more pronounced in these peripheral research domains (In Supplementary Table S8 and Figure 2a, $\beta = 0.059,\ p < 0.01$, representing only a 1.04-fold improvement for Physical sciences; compared to $\beta = 0.388,\ p < 0.01$, a 1.16-fold improvement for Health Sciences, and $\beta = 0.131,\ p < 0.01$, a 1.07-fold improvement for Life Sciences). Similarly, the coefficients for mere participation turn to significant and tend to be larger than those for shared leadership within medical-related studies (See Supplementary Table S9 and Figure 2, $\beta = 0.479,\ p < 0.01$ vs. $\beta = 0.347,\ p < 0.01$ for Health sciences; and $\beta = 0.178,\ p < 0.01$ vs. $\beta = 0.113,\ p < 0.01$ for Life sciences).

As illustrated in Figure 2b, the results remain consistent across approximately six decades, demonstrating that teams co-led by in-house scientists exhibit the highest probability of novelty relative to the baseline (an average 1.05- to 1.08-fold increases over the entire period, with the lowest confidence bound of 1.037). In contrast, the differences between teams with in-house scientists in a participatory role and the baseline remain consistently negligible (averaging 1.00-fold, 95%CI = [0.976, 1.027]).

### 5.2 Balanced user-staff team composition yields highest novelty return.

To test H3a and H3b, we transform the involvement of in-house scientists into two different continuous variables to explore potential non-linear dynamics and the risk of "cognitive lock-in". Building upon the fully specified model, Table 2 introduces quadratic terms to capture these non-linear effects. Specifically, we construct two continuous variables: the proportion of in-house scientists within the team (coded as 0 for teams without staff involvement), and the proportion of in-house scientists specifically within the team's leadership (coded as 0 for teams where staff merely participate without co-leading; note that external user only teams are excluded for this second setup).

In Table 2 column (1), the linear term for the proportion of in-house scientists is positive ($\beta = 0.399,\ p < 0.01$, while the squared term is significantly negative ($\beta = -0.434,\ p < 0.01$), confirming a distinct inverted U-shaped relationship. Figure 3a demonstrates that the relative probability of novelty peaks when in-house scientists constitute approximately 46% of the team (yielding an approximately 1.071-fold higher probability, 95%CI = [1.065, 1.076]) but it sharply loses its advantage and turns negative when their proportion exceeds 92%. Table 2 column (2) isolates the proportion of in-house scientists specifically within team leadership. A similar inverted U-shape emerges (linear: $\beta = 0.254,\ p < 0.05$; quadratic: $\beta = -0.397,\ p < 0.05$). Figure 3b shows that optimal novelty performance is achieved when in-house scientists hold about 30% of leadership positions (representing an approximately 1.032-fold improvement, 95%CI = [1.027, 1.035] times). However, if they dominate more than 60% of the leadership roles, the impact on novelty also becomes negative.

These results support H3 overall. They underscore that maximizing scientific novelty requires a balanced user-staff composition: disproportionate dominance by external users yields lower novelty returns due to insufficient epistemic support, while staff over-embeddedness could also stifle unconventional research through institutional routines.

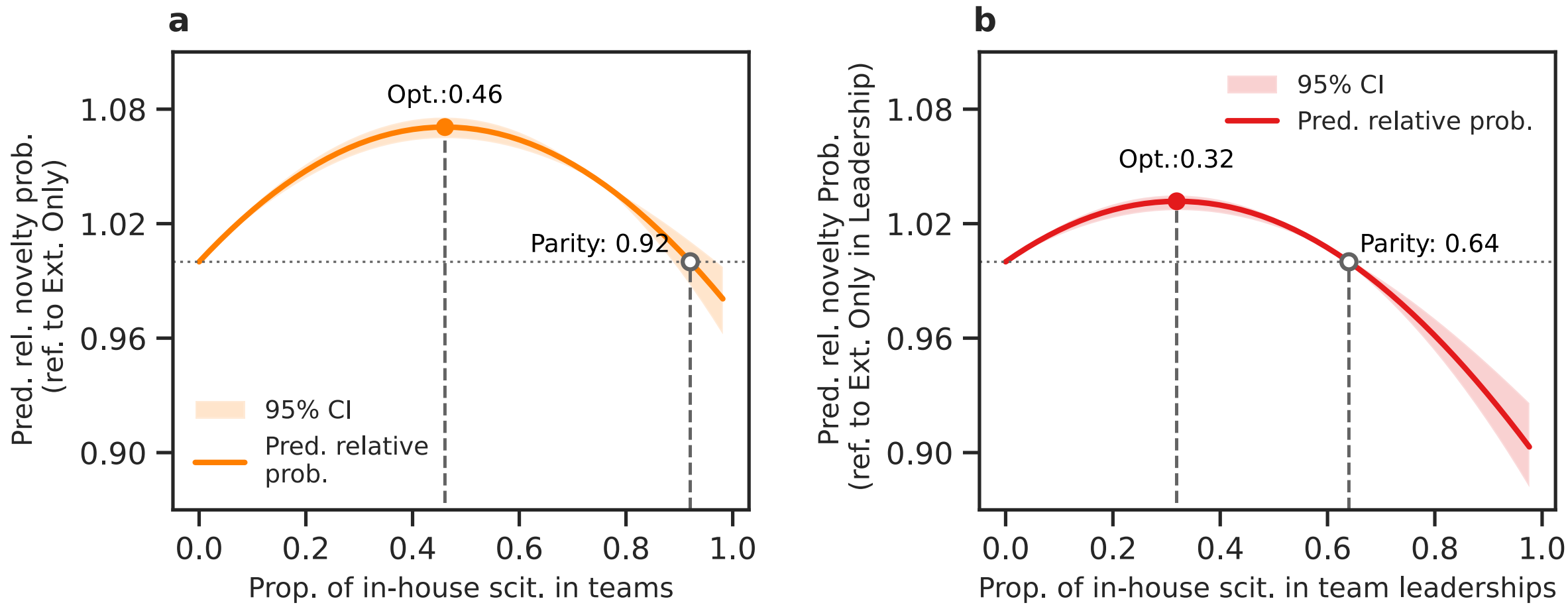

**Figure 3. Balanced user-staff team composition and leadership structure are associated with the highest novelty premium. a,** Predicted relative probability of scientific novelty as the proportion of in-house scientists in the team increases. Teams consisting purely of external users (proportion = 0) serve as the baseline. Predictions are based on the regression model shown in Table 2 column (1). The novelty premium peaks when in-house scientists constitute 46% of the team (approximately 7.1% higher, 95% CI = [1.065, 1.076]) and diminishes sharply when their proportion exceeds 92%. **b,** Predicted relative probability of scientific novelty as the proportion of in-house scientists in team leadership increases. The sample is restricted to user-staff teams containing at least one in-house scientist, with teams having zero in-house scientists in leadership role (proportion = 0) used as the reference group. Predictions are based on the model shown in Table 2 column (2). Optimal novelty performance (approximately 3.2% higher, 95% CI = [1.027, 1.035]) is achieved when in-house scientists hold about 32% of the leadership positions. The novelty effect turns negative when in-house scientists dominate more than 64% of the leadership roles. Error bars represent 95% standardized confidence intervals.

**Table 2. Logistic regression estimates indicating an inverted U-shaped relationship between the depth of in-house scientist involvement and scientific novelty.** All specifications include the full set of control variables and fixed effects for research discipline and publication year as defined in baseline models. Column (1) utilizes the overall proportion of in-house scientists within the team as the independent variable, while column (2) focuses on formal user-staff teams and uses as independent variable the proportion of in-house scientists within the team's leadership. The statistically significant negative quadratic terms in both regressions show that scientific novelty peaks at a relatively balanced collaborative equilibrium , suggesting the necessity of maintaining an epistemic balance between external users and staff scientists. Robust standard errors are reported in parentheses. Significance levels are based on two-sided Wald tests: *p<0.1, **p<0.05, ***p<0.01.

| | **DV: Novel paper = True** | |
|---|---|---|
| | (1) | (2) |
| **Prop. in-house scit. in team** | 0.399*** | |
| | (0.071) | |
| **Prop. in-house scit. in team$^2$** | -0.434*** | |
| | (0.122) | |
| **Prop. in-house scit. in leadership** | | 0.254** |
| | | (0.108) |
| **Prop. in-house scit.in leadership$^2$** | | -0.397** |
| | | (0.159) |
| **Num. author (log)** | -0.083*** | -0.287*** |
| | (0.007) | (0.015) |
| **International paper = True** | -0.047*** | -0.096*** |
| | (0.010) | (0.022) |
| **Num. references (log)** | 0.053*** | 0.099*** |
| | (0.009) | (0.018) |
| **Num. facilities** | 0.066*** | 0.106*** |
| | (0.007) | (0.011) |
| **SDG paper = True** | 0.084*** | 0.086*** |
| | (0.008) | (0.017) |
| **Avg. team age (log)** | 0.011 | -0.138*** |
| | (0.013) | (0.030) |
| **Avg. ext. leader avg. impact (log)** | -0.275*** | -0.189*** |
| | (0.007) | (0.015) |
| **Avg. ext. leader inst. h-index (log)** | -0.049*** | -0.044*** |
| | (0.008) | (0.015) |
| **Ext. leader Global North = True** | 0.309*** | 0.496*** |
| | (0.018) | (0.035) |
| **Ext. leader Same Country = True** | -0.060*** | -0.111*** |
| | (0.009) | (0.021) |
| **Ext. leader knowledge similarity** | -0.804*** | -0.905*** |
| | (0.160) | (0.322) |
| **Ext. leader avg. prior pub in Facility (log)** | -0.041*** | -0.001 |
| | (0.005) | (0.010) |
| **Ext. leader prior with in-house scit. = True** | 0.330*** | 0.059* |
| | (0.011) | (0.034) |
| ***Fixed-effects controls*** | | |
| **Disciplines (26 fields)** | Yes | Yes |
| **Publication year (51 years)** | Yes | Yes |
| **Obs.** | 294,879 | 78.522 |
| **Pseudo R$^2$** | 0.059 | 0.103 |

Beyond the main effects of collaboration patterns, the estimates of our control variables in Table 1 and Table 2 offer compelling independent insights into knowledge production within LSRIs. Notably, *Ext. leader knowledge similarity* consistently exhibits a highly significant negative coefficient (e.g., $\beta = -0.802,\ p < 0.01$ in Table 1 column (1)). This indicates that when external users' historical knowledge base strongly overlaps with the facility's prior outputs, the probability of producing novel knowledge significantly drops. Moreover, International papers are associated with lower novelty probabilities (e.g., $\beta = -0.047,\ p < 0.01$ in Table 1 column (2)), which aligns with findings by Wagner et al. (2019). Conversely, the *Ext. leader Global North* indicator is significantly positive (e.g., $\beta = 0.309,\ p < 0.01$ in Table 2 column (1)), highlighting persistent geographic and economic inequalities where teams led by researchers from developed nations are better positioned to extract novel outcomes from these capital-intensive infrastructures (Ke et al., 2026).

### 5.3 Gaining facility experience and the transition to deep collaboration architecture.

Table 3 and Figure 4 test H4a and H4b by introducing interaction terms to examine how external leaders' prior experience with the focal facility and its staff moderates the novelty premium. Overall, distinct collaboration patterns yield distinct returns, partly validating H4.

Table 3 column (1) shows that the interaction between external leaders' prior publications in the focal facility and in-house scientist participation is statistically negative ($\beta = -0.076,\ p < 0.01$). Figure 4a illustrates that as external users publish more at the focal facility, the relative novelty premium of formally having in-house scientists merely participate diminishes rapidly (the pink line). Notably, the decreasing trend in the co-leading team is insignificant ($\beta = 0.007,\ p = 0.497$), suggesting that shared leadership retains its relative value regardless of users' prior experience.

Table 3 column (2) explores the impact of prior collaboration with in-house scientists. The interaction terms between prior ties and in-house staff involvement are statistically negative for both participation ($\beta = -0.384,\ p < 0.01$) and co-leading ($\beta = -0.234,\ p < 0.01$). In contrast to our main regressions where staff participation yielded no significant benefits, the relative novelty probabilities shown in Figure 4b reveal a nuanced dynamic for new users: those without prior staff ties obtain a direct novelty premium from formal staff involvement, even in mere participatory roles. However, for seasoned users with pre-existing collaborative ties to facility staff, the effect of mere participation reverses, imposing a novelty penalty relative to teams purely composed of external users. This indicates that for experienced users, granting in-house scientists deeper, co-leading roles can help maintain more novel outcomes.

These findings partly validate H4 by revealing a clear divergence. For novice users, the novelty premium generated by mere staff participation is statistically comparable to that of co-leadership. However, the return on mere participation decreases sharply as external users accumulate experience; at this stage, sustaining novelty requires deeper, shared-leadership collaboration with staff scientists. This suggests that, as users mature in their infrastructure knowledge, team structures must evolve from mere infrastructural support toward genuine intellectual partnerships to overcome potential knowledge saturation and cognitive fatigue. Moreover, these results highlight a dynamic evolutionary pathway for user-staff collaborations: while integrating in-house scientists in either a participatory or co-leading capacity is effective for novices who are just bridging initial knowledge gaps, experienced users need complex, deeply integrated collaborative modes to sustain continued innovation.

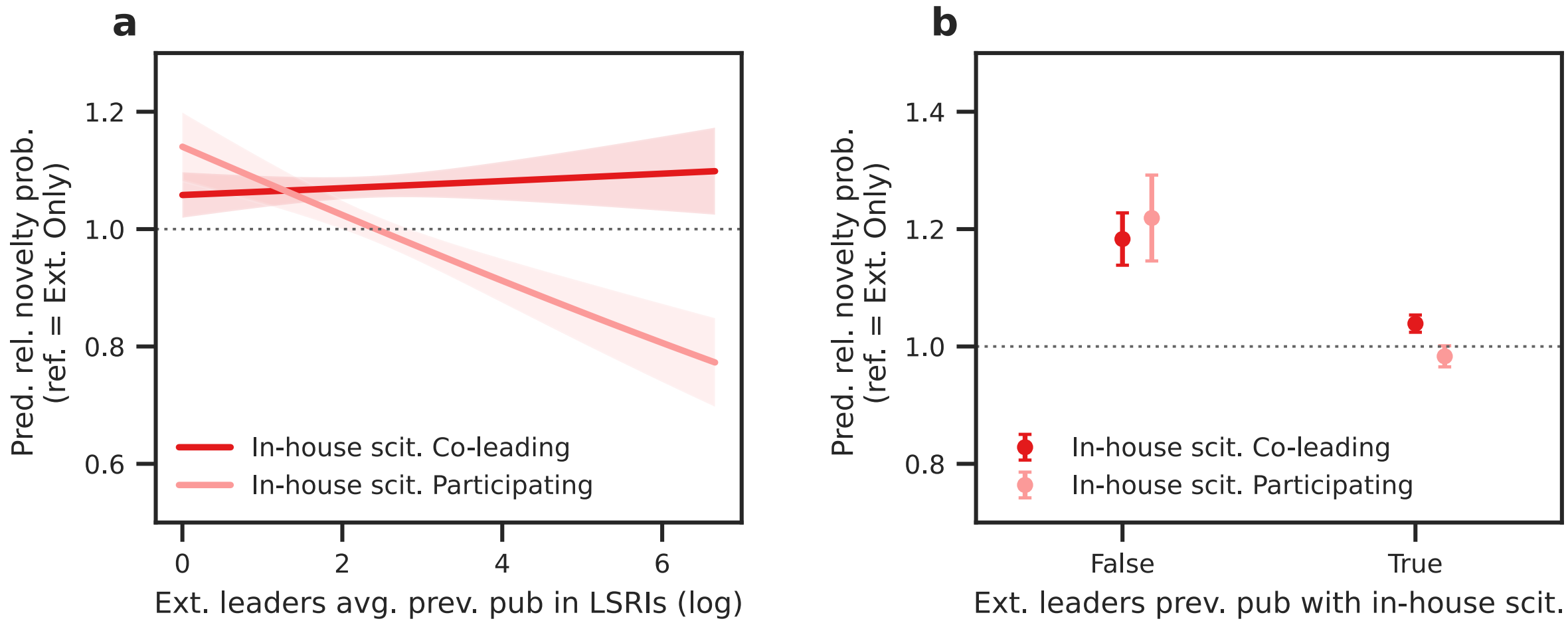

**Figure 4. Novice users benefit from mere staff participation, whereas experienced users need the deeper, co-leading collaboration architecture to sustain novelty. a,** Predicted relative probabilities of scientific novelty as a function of external leaders' historical publication volume at the focal facility. Teams consisting purely of external users are treated as the reference group. As external leaders publish more papers at the facility, the relative novelty premium associated with staff participation exhibits a sharp decline (pink line), whereas the benefit of shared co-leadership remains robust and stable (red line). Predictions are based on the regression model shown in Table 3 column (1). **b,** Predicted relative probabilities of scientific novelty as a function of external leaders' existing collaboration tie with in-house scientists, with teams consisting purely of external users treated as the reference group. For first-time collaborations, the novelty premium is statistically larger than for teams consisting only of external users, particularly when in-house scientists act in a participatory role (1.219, 95% CI = [1.146, 1.292]) vs. a co-leading role (1.183, 95% CI = [1.138, 1.228]). However, if external leaders possess prior ties with facility staff, the relative effect of mere participation turns negative (0.983, 95% CI = [0.965, 1.001]), while the co-leading collaboration architecture can help sustain the novelty premium (1.039, 95% CI = [1.024, 1.054]). Predictions are based on the model shown in Table 3 column (2); see Supplementary Table S10 and Supplementary Figure S8 for further evidence of diminishing returns. Error bars represent 95% standardized confidence intervals.

**Table 3. Logistic regression models with interaction terms show the diminishing return of user experience for both types of formal user-staff collaborations.** All specifications include the full set of control variables and fixed effects (research discipline and publication year) as defined in the baseline models. To estimate the moderating effect of user experience, we introduce two distinct interaction terms. Column (1) incorporates the external leaders accumulated prior publication volume at the focal facility as a continuous moderator to test for experience-driven diminishing returns. Column (2) employs a binary indicator of prior collaboration with in-house scientists to estimate the moderating effect of past collaboration ties. Robust standard errors are reported in parentheses. Significance levels are based on two-sided Wald tests: *p<0.1, **p<0.05, ***p<0.01.

| | **DV: Novel paper = True** | |
|---|---|---|
| | (1) | (2) |
| **In-house scit. participating = True** | 0.183*** | 0.356*** |
| | (0.036) | (0.055) |
| **In-house scit. co-leading = True** | 0.077*** | 0.299*** |
| | (0.025) | (0.032) |
| **Ext. leader avg. prior pub in Facility (log)** | -0.035*** | -0.036*** |
| | (0.006) | (0.005) |
| **In-house scit. participating X Ext. leader avg. prior pub in Facility (log)** | -0.076*** | |
| | (0.014) | |
| **In-house scit. co-leading X Ext. leader avg. prior pub in Facility (log)** | 0.007 | |
| | (0.010) | |
| **Ext. leader prior with in-house scit. = True** | 0.325*** | 0.358*** |
| | (0.011) | (0.012) |
| **In-house scit. participating X Ext. leader prior with in-house scit.** | | -0.384*** |
| | | (0.057) |
| **In-house scit. co-leading X Ext. leader prior with in-house scit.** | | -0.234*** |
| | | (0.034) |
| **Num. author (log)** | -0.078*** | -0.082*** |
| | (0.008) | (0.007) |
| **International paper = True** | -0.048*** | -0.049*** |
| | (0.010) | (0.010) |
| **Num. references (log)** | 0.053*** | 0.054*** |
| | (0.009) | (0.009) |
| **Num. facilities** | 0.067*** | 0.068*** |
| | (0.007) | (0.007) |
| **SDG paper = True** | 0.083*** | 0.084*** |
| | (0.008) | (0.008) |
| **Avg. team age (log)** | 0.008 | 0.007 |
| | (0.013) | (0.013) |
| **Avg. ext. leader avg. impact (log)** | -0.276*** | -0.274*** |
| | (0.007) | (0.007) |
| **Avg. ext. leader inst. h-index (log)** | -0.049*** | -0.050*** |
| | (0.008) | (0.008) |
| **Ext. leader Global North = True** | 0.310*** | 0.310*** |
| | (0.018) | (0.018) |
| **Ext. leader Same Country = True** | -0.062*** | -0.061*** |
| | (0.009) | (0.009) |
| **Ext. leader knowledge similarity** | -0.796*** | -0.788*** |
| | (0.160) | (0.160) |
| ***Fixed-effects controls*** | | |
| **Disciplines (26 fields)** | Yes | Yes |
| **Publication year (51 years)** | Yes | Yes |
| **Obs.** | 294,879 | 294,879 |
| **Pseudo $R^2$** | 0.060 | 0.060 |

To further unpack the negative moderating effect of historical user-staff ties (H4b), we conducted a nuanced exploration by replacing the binary prior formal collaboration indicator with two continuous variables: the accumulated number of prior formal collaborations where in-house scientists merely participated (Supplementary Table S10 column (1); Supplementary Figure S8a) and those where they shared leadership (Supplementary Table S10 column (2); Supplementary Figure S8b). According to Supplementary Table S10, the negative moderating effects exhibit distinct variations based on the nature of prior ties. When external leaders have a history of sharing leadership with in-house scientists, the returns on novelty significantly diminish regardless of whether current in-house scientists participate ($\beta = -0.183,\ p < 0.01$) or co-lead ($\beta = -0.211,\ p < 0.01$) the team, as visualized in Figure S8b. Conversely, an accumulation of mere participation ties exerts a drastically stronger negative moderating effect (see Figure S8a for visualization) on future staff participation ($\beta = -0.363,\ p < 0.01$) than on future co-leading ($\beta = -0.115,\ p < 0.01$).

These diverging distributions theoretically imply that the depth of historical knowledge transfer heavily dictates the value of future collaborations. Shared leadership inherently facilitates deeper epistemic integration and robust knowledge transfer; as a result, external users rapidly internalize the facility's tacit knowledge, leading to consistent diminishing returns across all future collaborative patterns. In contrast, mere participation offers limited knowledge transfer benefits to the users. Consequently, while repeated support-level participation quickly loses its novelty premium, granting in-house scientists' epistemic authority (co-leading) in subsequent projects can still effectively salvage and relatively sustain scientific novelty, as this deeper collaborative mode exhibits a higher tolerance for cognitive fatigue and knowledge saturation experienced by external users.

### 5.4 Robustness checks

To ensure the reliability of our findings, we conducted a series of robustness checks. First, we confirmed that all four hypotheses hold when incorporating facility-level fixed effects to account for unobserved, infrastructure-specific heterogeneities (See Table 4 for details). For simplicity, we report only the coefficients of the independent and moderating variables with all others confounding variables controlled. Column (1) and (2) present the results for H1 and H2, respectively, while the validations of H3 and H4 are shown from column (3) through (6). We maintain similar table structures for the subsequent robustness checks.

Second, recognizing the varied conventions of academic authorship, we re-estimated our models by replacing the aggregated characteristics of external leadership group with those of the first authors (Table 5) and last authors (Table 6). The core relationships—including the collaboration premium, the inverted U-shape, and the moderating effects of prior experience—remained highly consistent.

Finally, instead of treating scientific novelty as a binary outcome, we utilized the continuous novelty score as the dependent variable (Table 7), with positive values indicating conventionality and negative values indicating novelty. The regression results align with our main logistic regression models, confirming the structural validity of our theoretical framework.

**Table 4. Robustness checks incorporating facility-level fixed effects.** For brevity, only the coefficients of the independent and moderating variables are reported. Column (1) tests H1, while Column (2) evaluates H2. Columns (3) and (4) test the inverted U-shaped relationships proposed in H3a and H3b. Columns (5) and (6) estimate the moderating effects to verify the robustness of H4a and H4b. Standard errors, clustered at the facility-year level, are reported in parentheses. Significance levels are based on two-sided Wald tests: *p<0.1, **p<0.05, ***p<0.01.

| | DV: Novel paper = True | | | | | |
|---|---|---|---|---|---|---|
| | (1)<br>H1 | (2)<br>H2 | (3)<br>H3a | (4)<br>H3b | (5)<br>H4a | (6)<br>H4b |
| **With In-house scit. = True** | 0.071* | | | | | |
| | (0.043) | | | | | |
| **In-house scit. participating = True** | | 0.003 | | | 0.169** | 0.373*** |
| | | (0.046) | | | (0.085) | (0.055) |
| **In-house scit. co-leading = True** | | 0.103** | | | 0.134** | 0.305*** |
| | | (0.047) | | | (0.059) | (0.032) |
| **Prop. In-house scit. scientists in team** | | | 0.319 | | | |
| | | | (0.213) | | | |
| **Prop. In-house scit. scientists in team$^2$** | | | -0.170 | | | |
| | | | (0.244) | | | |
| **Prop. In-house scit. scientists in leadership** | | | | 0.467** | | |
| | | | | (0.198) | | |
| **Prop. In-house scit. scientists in leadership$^2$** | | | | -0.448* | | |
| | | | | (0.238) | | |
| **Ext. leader avg. prior pub in Facility (log)** | | | | | -0.049*** | |
| | | | | | (0.017) | |
| **In-h. participating X Ext. leader avg. prior pub in Facility (log)** | | | | | -0.070 | |
| | | | | | (0.030) | |
| **In-h. co-leading X Ext. leader avg. prior pub in Facility (log)** | | | | | -0.013 | |
| | | | | | (0.018) | |
| **Ext. leader prior with in-house scit. = True** | | | | | | 0.227*** |
| | | | | | | (0.026) |
| **In-h. participating X Ext. leader prior with in-house scit.** | | | | | | -0.287*** |
| | | | | | | (0.059) |
| **In-h. co-leading X Ext. leader prior with in-house scit.** | | | | | | -0.156*** |
| | | | | | | (0.047) |
| **Controls** | Yes | Yes | Yes | Yes | Yes | Yes |
| ***Fixed-effects controls*** | | | | | | |
| **Disciplines (26 fields)** | Yes | Yes | Yes | Yes | Yes | Yes |
| **Publication year (51 years)** | Yes | Yes | Yes | Yes | Yes | Yes |
| **Facility (74 facilities)** | Yes | Yes | Yes | Yes | Yes | Yes |
| **Obs.** | 294,851 | 294,851 | 294,851 | 78,508 | 294,851 | 294,851 |
| **Pseudo $R^2$** | 0.057 | 0.057 | 0.076 | 0.122 | 0.076 | 0.076 |

**Table 5. Robustness checks substituting external leader characteristics with those of the first authors.** For brevity, only the coefficients of the independent and moderating variables are reported. Column (1) tests H1, while Column (2) evaluates H2. Columns (3) and (4) test the inverted U-shaped relationships proposed in H3a and H3b. Columns (5) and (6) estimate the moderating effects to verify the robustness of H4a and H4b. Robust standard errors are reported in parentheses. Significance levels are based on two-sided Wald tests: *p<0.1, **p<0.05, ***p<0.01.

| | DV: Novel paper = True | | | | | |
|---|---|---|---|---|---|---|
| | (1)<br>H1 | (2)<br>H2 | (3)<br>H3a | (4)<br>H3b | (5)<br>H4a | (6)<br>H4b |
| **With In-house scit. = True** | 0.065***<br>(0.012) | | | | | |
| **In-house scit. participating = True** | | -0.006<br>(0.019) | | | 0.136***<br>(0.028) | 0.188***<br>(0.029) |
| **In-house scit. co-leading = True** | | 0.093***<br>(0.047) | | | 0.151***<br>(0.021) | 0.235***<br>(0.022) |
| **Prop. In-house scit. scientists in team** | | | 0.302***<br>(0.069) | | | |
| **Prop. In-house scit. scientists in team$^2$** | | | -0.237***<br>(0.082) | | | |
| **Prop. In-house scit. scientists in leadership** | | | | 0.249**<br>(0.098) | | |
| **Prop. In-house scit. scientists in leadership$^2$** | | | | -0.219**<br>(0.101) | | |
| **First author avg. prior pub in Facility (log)** | | | | | 0.021***<br>(0.006) | |
| **In-h. participating X First author avg. prior pub in Facility (log)** | | | | | -0.098***<br>(0.014) | |
| **In-h. co-leading X First author avg. prior pub in Facility (log)** | | | | | -0.038***<br>(0.010) | |
| **First author prior with in-house scit. = True** | | | | | | 0.251***<br>(0.015) |
| **In-h. participating X First author prior with in-house scit.** | | | | | | -0.345***<br>(0.036) |
| **In-h. co-leading X First author prior with in-house scit.** | | | | | | -0.244***<br>(0.027) |
| **Controls** | Yes | Yes | Yes | Yes | Yes | Yes |
| ***Fixed-effects controls*** | | | | | | |
| **Disciplines (26 fields)** | Yes | Yes | Yes | Yes | Yes | Yes |
| **Publication year (51 years)** | Yes | Yes | Yes | Yes | Yes | Yes |
| **Obs.** | 209,290 | 209,290 | 209,290 | 62,738 | 209,290 | 209,290 |
| **Pseudo $R^2$** | 0.057 | 0.057 | 0.076 | 0.098 | 0.057 | 0.058 |

**Table 6. Robustness checks substituting external leader characteristics with those of the last authors.** For brevity, only the coefficients of the independent and moderating variables are reported. Column (1) tests H1, while Column (2) evaluates H2. Columns (3) and (4) test the inverted U-shaped relationships proposed in H3a and H3b. Columns (5) and (6) estimate the moderating effects to verify the robustness of H4a and H4b. Robust standard errors are reported in parentheses. Significance levels are based on two-sided Wald tests: *p<0.1, **p<0.05, ***p<0.01.

| | DV: Novel paper = True | | | | | |
|---|---|---|---|---|---|---|
| | (1)<br>H1 | (2)<br>H2 | (3)<br>H3a | (4)<br>H3b | (5)<br>H4a | (6)<br>H4b |
| **With In-house scit. = True** | 0.089***<br>(0.010) | | | | | |
| **In-house scit. participating = True** | | 0.015<br>(0.016) | | | 0.144***<br>(0.033) | 0.313***<br>(0.035) |
| **In-house scit. co-leading = True** | | 0.120***<br>(0.011) | | | 0.090***<br>(0.025) | 0.264***<br>(0.027) |
| **Prop. In-house scit. scientists in team** | | | 0.345***<br>(0.059) | | | |
| **Prop. In-house scit. scientists in team$^2$** | | | -0.191***<br>(0.071) | | | |
| **Prop. In-house scit. scientists in leadership** | | | | 0.146*<br>(0.086) | | |
| **Prop. In-house scit. scientists in leadership$^2$** | | | | -0.108<br>(0.088) | | |
| **Last author avg. prior pub in Facility (log)** | | | | | -0.048***<br>(0.005) | |
| **In-h. participating X Last author avg. prior pub in Facility (log)** | | | | | -0.049***<br>(0.011) | |
| **In-h. co-leading X Last author avg. prior pub in Facility (log)** | | | | | 0.010<br>(0.008) | |
| **Last author prior with in-house scit. = True** | | | | | | 0.253***<br>(0.012) |
| **In-h. participating X Last author prior with in-house scit.** | | | | | | -0.183***<br>(0.029) |
| **In-h. co-leading X Last author prior with in-house scit.** | | | | | | -0.375***<br>(0.038) |
| **Controls** | Yes | Yes | Yes | Yes | Yes | Yes |
| ***Fixed-effects controls*** | | | | | | |
| **Disciplines (26 fields)** | Yes | Yes | Yes | Yes | Yes | Yes |
| **Publication year (51 years)** | Yes | Yes | Yes | Yes | Yes | Yes |
| **Obs.** | 273,125 | 273,125 | 273,125 | 79,026 | 273,125 | 273,125 |
| **Pseudo $R^2$** | 0.056 | 0.056 | 0.056 | 0.095 | 0.056 | 0.056 |

**Table 7. Robustness checks using the continuous novelty score as the dependent variable.** For brevity, only the coefficients of the independent and moderating variables are reported. Column (1) tests H1, while Column (2) evaluates H2. Columns (3) and (4) test the inverted U-shaped relationships proposed in H3a and H3b. Columns (5) and (6) estimate the moderating effects to verify the robustness of H4a and H4b. Robust standard errors are reported in parentheses. Significance levels are based on two-sided Wald tests: *p<0.1, **p<0.05, ***p<0.01.

| | DV: Novelty score | | | | | |
|---|---|---|---|---|---|---|
| | (1)<br>H1 | (2)<br>H2 | (3)<br>H3a | (4)<br>H3b | (5)<br>H4a | (6)<br>H4b |
| **With In-h. scientists = True** | -7.750***<br>(0.648) | | | | | |
| **In-h. participating = True** | | -2.470**<br>(1.080) | | | -38.500***<br>(3.180) | -26.6***<br>(3.230) |
| **In-h. co-leading = True** | | -10.2***<br>(0.653) | | | -6.050***<br>(1.990) | -23.5***<br>(2.300) |
| **Prop. in-h. scientists in team** | | | -50.9***<br>(4.70) | | | |
| **Prop. in-h. scientists in team$^2$** | | | 47.2***<br>(8.76) | | | |
| **Prop. in-h. scientists in leadership** | | | | 11.9*<br>(6.76) | | |
| **Prop. in-h. scientists in leadership$^2$** | | | | -6.99<br>(9.18) | | |
| **Ext. leader avg. prior pub in Facility (log)** | | | | | 0.408<br>(0.512) | |
| **In-h. participating X Ext. leader avg. prior pub in Facility (log)** | | | | | 14.700***<br>(1.090) | |
| **In-h. co-leading X Ext. leader avg. prior pub in Facility (log)** | | | | | -1.760**<br>(0.691) | |
| **Ext. leader avg. prior with in-house scit. = True** | | | | | | -30.100***<br>(1.020) |
| **In-h. participating X Ext. leader avg. prior with in-house scit.** | | | | | | 26.100***<br>(3.390) |
| **In-h. co-leading X Ext. leader avg. prior with in-house scit.** | | | | | | 15.100***<br>(2.340) |
| **Controls** | Yes | Yes | Yes | Yes | Yes | Yes |
| ***Fixed-effects controls*** | | | | | | |
| **Disciplines (26 fields)** | Yes | Yes | Yes | Yes | Yes | Yes |
| **Publication year (51 years)** | Yes | Yes | Yes | Yes | Yes | Yes |
| **Obs.** | 294,781 | 294,781 | 294,781 | 78,519 | 294,781 | 294,781 |
| **Pseudo $R^2$** | 0.060 | 0.060 | 0.060 | 0.116 | 0.061 | 0.061 |

## 6 Discussion

### 6.1 Theoretical implications

This study makes several contributions to the literature on facilitymetrics and the team dynamics of knowledge production in large-scale research infrastructures.

First, we advance the theoretical framework of user-staff collaboration conceptualized by D'Ippolito and Rüling (2019). While their seminal interview-based study at a single neutron facility identified diverse collaborative archetypes, ranging from “full service” to “peer collaboration”, based on qualitative perceptions of expertise gaps, our study provides the first large-scale empirical validation of these archetypes. By leveraging a validated machine-learning classification framework for author contributions from F. L. Xu et al. (2022) and using affiliations to identify in-house scientists, we successfully mapped their qualitative typologies onto quantitative collaboration patterns (*Only external users*, *In-house scientists participating*, and *In-house scientists co-leading*). Crucially, we shift the theoretical focus from the formation of these collaborations to their epistemological consequences. We demonstrate that treating the collaborations with in-house scientists informally or merely as supporters in their teams preserves an asymmetrical power structure that fails to maximize the facility's innovative potential. Instead, granting them epistemic authority (co-leadership) flattens the team's micro-power structure (H. M. Xu et al., 2022; Xu et al., 2024) and is associated with high novelty probabilities. This symmetrical distribution of epistemic authority empowers in-house scientists to proactively utilize the facilities’ capabilities, reshape users’ research frameworks, and foster genuine cognitive diversity and knowledge recombination without the hierarchical suppression often found in traditional service provider dynamics (Lauto & Valentin, 2013; Mansoor et al., 2013).

Second, we enrich the literature on facilitymetrics and the division of scientific labor by quantifying the optimal team composition of user-staff collaboration. While qualitative accounts suggest that structural integration between in-house scientists and users is beneficial, our non-linear findings introduce a critical theoretical caveat: the dual risks of collaborative imbalance and its downstream “epistemic lock-in” (Liu et al., 2022). We establish that while the boundary-spanning capabilities of in-house scientists bridge the gap between the limits of huge machines and frontier science, their involvement must be carefully bounded and balance-oriented. Over-reliance on internal staff risks introducing routine knowledge from their backgrounds and subordinating the users' frontier scientific questions to established institutional conventions (Zhang, Wang, et al., 2025). Conversely, avoiding this does not mean reverting to the absolute dominance of external users; rather, optimizing scientific novelty requires a delicate epistemic equilibrium where neither party monopolizes the research agenda. Such a balance structure might ensure that external frontier inquiries are neither hindered by an under-utilization of the cutting-edge facilities, nor stifled by its institutional routines. This quantitatively not only substantiates the necessity of the “user-oriented” model in modern Big Science (Hallonsten & Christensson, 2017; Lauto & Valentin, 2013), but also demonstrates that it is best maintained not through user isolation, but through balanced collaborative partnerships.

Finally, our study contributes to the dynamics of scientific collaboration networks by empirically capturing the lifecycle of user-staff ties. D'Ippolito and Rüling (2019) noted that users' needs evolve as they acquire experiential knowledge. We formalize this dynamic by revealing an evolutionary shift in collaborative needs rather than a simple narrative of diminishing returns. We establish a dynamic evolution in knowledge transfer: for novice users and those collaborating with staff for the first time, foundational technical participation yields novelty premium comparable to those of deep epistemic integration (co-leading). Given that attempting deep integration prematurely might impose higher coordination costs without delivering proportional gains, these newcomers are adequately served by the low-friction support of mere staff participation. However, as external users mature and bridge the initial knowledge gap, this participatory premium depreciates due to potential knowledge saturation (Liu et al., 2024; Santos et al., 2024). In contrast, highly experienced users face a divergent path: they require genuine intellectual partnerships with staff scientists (co-leadership rather than mere support) to maintain creative exploitation of the same facility. By highlighting that the optimal division of scientific labor in LSRIs is not static but dynamically evolves with the accumulation of user experience (Langford & Langford, 2000). our study pushes the boundary of Social Exchange Theory within the context of machine-driven scientific research.

### 6.2 Policy and managerial implications

Given the intense competition among external users for limited experimental beamtime, facilities must optimize their human capital to maximize scientific returns (D'Ippolito & Rüling, 2019; Hallonsten, 2011). In this context, our empirical evidence provides critical policy implications for the management of LSRIs.

First, facilities and funding agencies (e.g., DOE, NSF, ERC, NSFC) must transition from a traditional service-oriented paradigm to one that actively incentivizes structural user-staff integration. Institutional evaluation metrics should be reformed to formally reward the "invisible labor" and co-leading contributions of in-house scientists, rather than evaluating them solely on technical maintenance.

Second, maximizing scientific novelty requires a flexible intervention model that carefully calibrates staff formal or informal involvement based on users' evolving collaborative needs. We recommend a tiered user-support and funding system tailored to experiential knowledge. For novice users and first-time collaborators, targeted grants could actively encourage in-house scientists to assume participatory roles. Given that the premium of staff participation and co-leadership are statistically comparable for new users, this encouragement shall maximize the limited time and energy of these staff across multiple user teams and participatory approach can effectively lower initial knowledge gaps without the coordination frictions that typically arise from premature deep integration (D'Ippolito & Rüling, 2019). Conversely, for highly experienced users, mere participation risks stifling novelty due to potential knowledge saturation. For these seasoned facility users, management should foster genuine shared-leadership partnerships to enable deep epistemic integration or at least grant greater research autonomy to push the machine's absolute boundaries.

### 6.3 Limitations and future research

Despite the large-scale data and hybrid machine-learning classification approach, this study has several limitations. First, our identification of team roles relies on post-hoc bibliometric data (author contribution statements and byline sequences). Beyond the bounded predictive accuracy of our machine-learning classifier, inferring roles exclusively from authorship records is inherently limited because collaboration does not strictly equate to co-authorship. This approach cannot capture "invisible labor" where in-house scientists provide critical support but are denied formal authorship (Lu et al., 2020; Sauermann & Haeussler, 2017), potentially overestimating the novelty performance of teams consisting purely of external users and underestimating the effect sizes of our findings. Moreover, building on post-hoc data also introduces self-selection bias and endogeneity to our research, even though we employed robust Propensity Score Matching (PSM) and fixed-effects models to mitigate confounding factors. It is plausible that external user proposals exhibiting exceptional novelty inherently attract in-house scientists to assume co-leading roles, rather than the collaborative structure uniquely driving the novelty.

Second, our reliance on institutional affiliation matching (ROR IDs) via OpenAlex to identify in-house scientists is susceptible to data truncation. This approach inevitably omits facility staff lacking standardized affiliation metadata in OpenAlex or those who transition between academia and facilities without promptly updating their affiliations in their publications (L. Zhang et al., 2024). Third, our data curation strategy, which aggregates records from the self-constructed publication databases maintained by various facilities, inherently carries risks of data loss, time lags, and reporting inconsistencies (Zhang, Wang, et al., 2026). Furthermore, due to the uneven levels of data openness globally, our sample was restricted to facilities maintaining mature, accessible publication lists that met our specific criteria. Therefore, our findings may not fully generalize to smaller or less transparent user-oriented facilities omitted from our list.

Future research could utilize qualitative methodologies, such as ethnographic studies or structured interviews at specific facilities, to explore the micro-sociological mechanisms of how in-house scientists negotiate epistemic authority with external leaders. From a quantitative perspective, future studies could employ Large Language Models (LLMs) to mine the acknowledgment sections of publications, thereby capturing and quantifying the informal, non-authorship contributions of in-house scientists (Danús et al., 2026). Additionally, leveraging quasi-experimental designs, such as Difference-in-Differences (DiD) estimators or instrumental variables (IVs), could help establish stronger causal mechanisms between the depth of user-staff integration and the trajectory of scientific novelty.

## 7 Conclusion

As large-scale research infrastructures increasingly push the frontiers of modern science, the collaborative dynamics between external users who pose scientific questions and in-house scientists who operate these huge machines are becoming critically important to understand. Drawing on a comprehensive dataset of 273,109 publications from 76 global Big Science facilities and employing advanced machine-learning techniques to parse the division of scientific labor, this study provides a systematic empirical investigation of the novelty premium of different types of user-staff collaborations. We find that formally incorporating in-house scientists into external teams as co-authors yields a significant novelty benefit, especially when they are granted shared leadership rather than relegated to mere participation or technical service roles. Crucially, to avoid the dual pitfalls of epistemic lock-in, user-staff collaboration should be dynamically managed: collaborative teams need to strike a delicate epistemic equilibrium where neither party monopolizes the research agenda, and the nature of in-house scientist involvement should adapt as users transition from novice to experienced. Ultimately, advancing the frontier of Big Science requires not only powerful machines but also the structural alignment between the diverse minds that operate them.

## Data and code availability

The datasets generated and analyzed for the current study are available in an openly accessible Github repository at: https://github.com/zhangmingze-ss/BSF_collaboration. The publication list of global LSRIs curated by the authors are available from the corresponding author upon reasonable request. The OpenAlex database used in this study could be accessed at: https://docs.openalex.org/.

## Acknowledgments

This work was supported by the National Social Science Fund Major Projects of China (Project No. 22&ZD127) and was partially supported by the City University of Hong Kong Startup Grant (Project No. 9610703). We would like to thank Yuhui DONG, Honghong LI, and Ren WEI for their expertise and assistances with the basic knowledge of large-scale research infrastructures. Many thanks to Meijun LIU, Yi BU, Zhesi SHEN, and Lingling ZHANG for their insightful comments on our earlier drafts. We would like to thank constructive comments from the journal editor and external reviewers.

**Author contribution statement**

Mingze ZHANG: Data curation, Formal analysis, Visualization, Writing – original draft; Yizhan LI: Conceptualization, Writing – review & editing; Hao PENG: Conceptualization, Validation, Writing – review & editing, Supervision; Zexia LI: Funding acquisition, Project administration, Validation, Writing – review & editing.

**Competing interests**

The authors declare no competing interests.

**Declaration of AI usage**

The authors declare that the Gemini-3.1 Pro was used to polish the language for readability and Deepseek-R1 was used to classify team roles when training the machine-learning classifier (See 4.2 Measurements). The authors have not used this tool in any other capacity, such as for idea generation, conceptualization, data analysis, generation of insights, or visualization. After using LLM, the authors reviewed and edited the content and take full responsibility for the content of the published article.

**References**

Anupama, G., Chattopadhyay, S., Deshpande, S., Ghosh, J., M. Godbole, R., Indumathi, D., & Souradeep, T. (2021). Big science in India. *Nature Reviews Physics*, *3*(11), 728-731.

Bianco, W., Gerhart, D., & Nicolson-Crotty, S. (2017). Waypoints for Evaluating Big Science* [Article]. *Social Science Quarterly*, *98*(4), 1144-1150. https://doi.org/10.1111/ssqu.12467

Blau, P. (2017). *Exchange and power in social life*. Routledge.

Börner, K., Silva, F. N., & Milojevic, S. (2021). Visualizing big science projects [Review]. *Nature Reviews Physics*, *3*(11), 753-761. https://doi.org/10.1038/s42254-021-00374-7

Brumfiel, G. (2011). The collider that cried 'Higgs'. *Nature*, *473*(7346), 136-137.

Castelvecchi, D. (2017). Gravitational wave detection wins physics Nobel. *Nature*, *550*(7674), NA-NA.

Cheng, D., Xue, Z., Zhibo, Y., & Mingze, Z. (2025). Impact of interdisciplinarity on disruptive innovation: the moderating role of collaboration pattern and collaboration size. *Scientometrics*. https://doi.org/10.1007/s11192-025-05285-3

Cherezov, V., Rosenbaum, D. M., Hanson, M. A., Rasmussen, S. G., Thian, F. S., Kobilka, T. S., Choi, H.-J., Kuhn, P., Weis, W. I., & Kobilka, B. K. (2007). High-resolution crystal structure of an engineered human β2-adrenergic G protein–coupled receptor. *Science*, *318*(5854), 1258-1265.

Cohan, A., Feldman, S., Beltagy, I., Downey, D., & Weld, D. S. (2020). Specter: Document-level representation learning using citation-informed transformers. Proceedings of the 58th annual meeting of the association for computational linguistics,

Conroy, G. (2024). World's brightest X-rays: China first in Asia to build next-generation synchrotron. *Nature*, *629*(8013), 740. https://doi.org/10.1038/d41586-024-01346-4

Crease, R. P., & Westfall, C. (2016). The new big science. *Physics Today*, *69*(5), 30-36.

Cropanzano, R., & Mitchell, M. S. (2005). Social exchange theory: An interdisciplinary review. *Journal of Management*, *31*(6), 874-900.

D'Ippolito, B., & Rüling, C. C. (2019). Research collaboration in Large Scale Research Infrastructures: Collaboration types and policy implications [Article]. *Research Policy*, *48*(5), 1282-1296. https://doi.org/10.1016/j.respol.2019.01.011

Danús, L., Dinneen, W., Torreblanca, C., Grossman, G., & González-Bailón, S. (2026). Informal connections outweigh coauthorship ties in academic impact. *Proceedings of the National Academy of Sciences*, *123*(18), e2511050123. https://doi.org/doi:10.1073/pnas.2511050123

Duan, Y. R., Memon, S. A., Alshebli, B., Guan, Q., Holme, P., & Lrahwan, T. (2025). Postdoc publications and citations link to academic retention and faculty success [Article]. *Proceedings of the National Academy of Sciences of the United States of America*, *122*(4), 3, Article e2402053122. https://doi.org/10.1073/pnas.2402053122

Giffoni, F., & Florio, M. (2023). Public support of science: A contingent valuation study of citizens' attitudes about CERN with and without information about implicit taxes [Article]. *Research Policy*, *52*(1), 19, Article 104627. https://doi.org/10.1016/j.respol.2022.104627

Guimerà, R., Uzzi, B., Spiro, J., & Amaral, L. A. N. (2005). Team assembly mechanisms determine collaboration network structure and team performance [Article]. *Science*, *308*(5722), 697-702. https://doi.org/10.1126/science.1106340

Hallonsten, O. (2011). Growing big science in a small country: MAX-lab and the Swedish research policy system. *Historical Studies in the Natural Sciences*, *41*(2), 179-215.

Hallonsten, O. (2013). Introducing 'facilitymetrics': a first review and analysis of commonly used measures of scientific leadership among synchrotron radiation facilities worldwide [Article]. *Scientometrics*, *96*(2), 497-513. https://doi.org/10.1007/s11192-012-0945-9

Hallonsten, O. (2014). How expensive is Big Science? Consequences of using simple publication counts in performance assessment of large scientific facilities [Article]. *Scientometrics*, *100*(2), 483-496. https://doi.org/10.1007/s11192-014-1249-z

Hallonsten, O. (2016a). *Big Science Transformed: Science, Politics and Organization in Europe and the United States*. Springer International Publishing. https://books.google.com.hk/books?id=fSpEDQAAQBAJ

Hallonsten, O. (2016b). Use and productivity of contemporary, multidisciplinary Big Science [Article]. *Research Evaluation*, *25*(4), 486-495. https://doi.org/10.1093/reseval/rvw019

Hallonsten, O., & Christensson, O. (2017). Collaborative technological innovation in an academic, uaser-oriented Big Science facility [Article]. *Industry and Higher Education*, *31*(6), 399-408. https://doi.org/10.1177/0950422217729284

Hao, Q., Xu, F., Li, Y., & Evans, J. (2026). Artificial intelligence tools expand scientists' impact but contract science's focus. *Nature*, 1-7.

Heidler, R., & Hallonsten, O. (2015). Qualifying the performance evaluation of Big Science beyond productivity, impact and costs [Article]. *Scientometrics*, *104*(1), 295-312. https://doi.org/10.1007/s11192-015-1577-7

Heinze, T., & Hallonsten, O. (2017). The reinvention of the SLAC National Accelerator Laboratory, 1992-2012 [Article]. *History and Technology*, *33*(3), 300-332. https://doi.org/10.1080/07341512.2018.1449711

Hoekman, J., & Rake, B. (2024). Geography of authorship: How geography shapes authorship attribution in big team science. *Research Policy*, *53*(2), 104927. https://doi.org/https://doi.org/10.1016/j.respol.2023.104927

Hofstra, B., Kulkarni, V. V., Munoz-Najar Galvez, S., He, B., Jurafsky, D., & McFarland, D. A. (2020). The diversity–innovation paradox in science. *Proceedings of the National Academy of Sciences*, *117*(17), 9284-9291.

Hoogendoorn, S., Oosterbeek, H., & Van Praag, M. (2012). Ethnic diversity and team performance: a randomized field experiment. Academy of Management Proceedings,

Jiménez, C. (2010). Synching Europe's big science facilities. *Nature*, - *464*(- 7289), - 659. 10.1038/464659a

Jones, B. F., Wuchty, S., & Uzzi, B. (2008). Multi-University Research Teams: Shifting Impact, Geography, and Stratification in Science [Article]. *Science*, *322*(5905), 1259-1262. https://doi.org/10.1126/science.1158357

Ke, Q., Pan, T., & Mao, J. (2026). The geography of novel and atypical research. *Research Policy*, *55*(1), 105345. https://doi.org/https://doi.org/10.1016/j.respol.2025.105345

Kozheurov, Y., & Teimurov, E. (2020). Megascience facilities in global research infrastructure. *Journal of Physics: Conference Series*, *1685*(1), 012014. https://doi.org/10.1088/1742-6596/1685/1/012014

Langford, C. H., & Langford, M. W. (2000). The evolution of rules for access to megascience research environments viewed from Canadian experience. *Research Policy*, *29*(2), 169-179.

Lauto, G., & Valentin, F. (2013). How Large-Scale Research Facilities Connect to Global Research [Article]. *Review of Policy Research*, *30*(4), 381-408. https://doi.org/10.1111/ropr.12027

Li, W., Zheng, H., Brand, J. E., & Clauset, A. (2025). Gender and racial diversity socialization in science. *Nature Computational Science*, *5*(6), 481-491.

Lin, Y., Frey, C. B., & Wu, L. (2023). Remote collaboration fuses fewer breakthrough ideas. *Nature*, *623*(7989), 987-991. https://doi.org/10.1038/s41586-023-06767-1

Lin, Z., Yin, Y., Liu, L., & Wang, D. (2023). SciSciNet: A large-scale open data lake for the science of science research. *Scientific Data*, *10*(1), 315. https://doi.org/10.1038/s41597-023-02198-9

Liu, J. W., Guo, X. F., Xu, S., Bu, Y., Sugimoto, C. R., Lariviere, V., Song, Y. L., & Zhou, H. H. (2024). Understanding super-partnerships in scientific collaboration: Evidence from the field of economics [Article]. *Journal of the Association for Information Science and Technology*, *75*(6), 717-733. https://doi.org/10.1002/asi.24876

Liu, M. J., Jaiswal, A., Bu, Y., Min, C., Yang, S. J., Liu, Z. B., Acuna, D., & Ding, Y. (2022). Team formation and team impact: The balance between team freshness and repeat collaboration [Article]. *Journal of Informetrics*, *16*(4), 15, Article 101337. https://doi.org/10.1016/j.joi.2022.101337

Lu, C., Zhang, C. W., Xiao, C. R., & Ding, Y. (2022). Contributorship in scientific collaborations: The perspective of contribution-based byline orders [Article]. *Information Processing & Management*, *59*(3), 14, Article 102944. https://doi.org/10.1016/j.ipm.2022.102944

Lu, C., Zhang, Y. Y., Ahn, Y. Y., Ding, Y., Zhang, C. W., & Ma, D. D. (2020). Co-contributorship Network and Division of Labor in Individual Scientific Collaborations [Article]. *Journal of the Association for Information Science and Technology*, *71*(10), 1162-1178. https://doi.org/10.1002/asi.24321

Malerba, F., & Orsenigo, L. (1995). Schumpeterian patterns of innovation. *Cambridge Journal of Economics*, *19*(1), 47-65. http://www.jstor.org/stable/23599565

Manganote, E. J. T., Schulz, P. A., & Cruz, C. H. D. (2016). Effect of high energy physics large collaborations on higher education institutions citations and rankings [Article]. *Scientometrics*, *109*(2), 813-826. https://doi.org/10.1007/s11192-016-2048-5

Mansoor, H. S., Ali, H., Ali, N., & Ali, H. (2013). Cognitive diversity and team performance: a review. *Journal of Basic and Applied Scientific Research*, *3*(6), 9-13.

Mello, A. L., & Rentsch, J. R. (2015). Cognitive diversity in teams: A multidisciplinary review. *Small Group Research*, *46*(6), 623-658.

Montoya, R. M., & Horton, R. S. (2013). A meta-analytic investigation of the processes underlying the similarity-attraction effect. *Journal of social and personal relationships*, *30*(1), 64-94.

Nogrady, B. (2023). HYPERAUTHORSHIP AND WHAT IT MEANS FOR 'BIG TEAM' SCIENCE [Editorial Material]. *Nature*, *615*(7950), 175-177. https://doi.org/10.1038/d41586-023-00575-3

Paletz, S. B., Peng, K., Erez, M., & Maslach, C. (2004). Ethnic composition and its differential impact on group processes in diverse teams. *Small Group Research*, *35*(2), 128-157.

Peng, H., Teplitskiy, M., & Jurgens, D. (2024). Author mentions in science news reveal widespread disparities across name-inferred ethnicities. *Quantitative Science Studies*, *5*(2), 351-365. https://doi.org/10.1162/qss_a_00297

Price, D. J. D. S. (1963). Columbia University Press. https://doi.org/doi:10.7312/pric91844

Priem, J., Piwowar, H. A., & Orr, R. (2022). OpenAlex: A fully-open index of scholarly works, authors, venues, institutions, and concepts. *ArXiv*, *abs/2205.01833*.

Rasmussen, S. G. F., DeVree, B. T., Zou, Y., Kruse, A. C., Chung, K. Y., Kobilka, T. S., Thian, F. S., Chae, P. S., Pardon, E., Calinski, D., Mathiesen, J. M., Shah, S. T. A., Lyons, J. A., Caffrey, M., Gellman, S. H., Steyaert, J., Skiniotis, G., Weis, W. I., Sunahara, R. K., & Kobilka, B. K. (2011). Crystal structure of the β2 adrenergic receptor–Gs protein complex. *Nature*, *477*(7366), 549-555. https://doi.org/10.1038/nature10361

Santos, J. M., Horta, H., & Feng, S. (2024). Homophily and its effects on collaborations and repeated collaborations: a study across scientific fields. *Scientometrics*, *129*(3), 1801-1823. https://doi.org/10.1007/s11192-024-04950-3

Sauermann, H., & Haeussler, C. (2017). Authorship and contribution disclosures. *Science Advances*, *3*(11), e1700404.
Seo, W., & Bu, Y. (2025). Transforming role classification in scientific teams using LLMs and advanced predictive analytics. *Quantitative Science Studies*, *6*, 505-523. https://doi.org/10.1162/qss_a_00360
Silva, F. S. V., Schulz, P. A., & Noyons, E. C. M. (2019). Co-authorship networks and research impact in large research facilities: benchmarking internal reports and bibliometric databases [Article]. *Scientometrics*, *118*(1), 93-108. https://doi.org/10.1007/s11192-018-2967-4
Söderström, K. R. (2023a). Global reach, regional strength: Spatial patterns of a big science facility [Article]. *Journal of the Association for Information Science and Technology*, *74*(9), 1140-1156. https://doi.org/10.1002/asi.24811
Söderström, K. R. (2023b). The structure and dynamics of instrument collaboration networks [Article]. *Scientometrics*, *128*(6), 3581-3600. https://doi.org/10.1007/s11192-023-04658-w
Söderström, K. R., Åström, F., & Hallonsten, O. (2022). Generic instruments in a synchrotron radiation facility [Article]. *Quantitative Science Studies*, *3*(2), 420-442. https://doi.org/10.1162/qss_a_00190
Stix, G. (2001). Little big science. *Scientific American*, *285*(3), 32-37.
Teplitskiy, M., Peng, H., Blasco, A., & Lakhani, K. R. (2022). Is novel research worth doing? Evidence from peer review at 49 journals. *Proceedings of the National Academy of Sciences*, *119*(47), e2118046119.
Thelwall, M., Kousha, K., Abdoli, M., Stuart, E., Makita, M., Wilson, P., & Levitt, J. (2024). Which international co-authorships produce higher quality journal articles? [Article; Early Access]. *Journal of the Association for Information Science and Technology*, 20. https://doi.org/10.1002/asi.24881
Tian, C., Huang, Y., Jin, C., Ma, Y., & Uzzi, B. (2025). The distinctive innovation patterns and network embeddedness of scientific prizewinners. *Proceedings of the National Academy of Sciences*, *122*(40), e2424143122. https://doi.org/doi:10.1073/pnas.2424143122
Tscharntke, T., Hochberg, M. E., Rand, T. A., Resh, V. H., & Krauss, J. (2007). Author Sequence and Credit for Contributions in Multiauthored Publications. *Plos Biology*, *5*(1), e18. https://doi.org/10.1371/journal.pbio.0050018
Uzzi, B., Mukherjee, S., Stringer, M., & Jones, B. (2013). Atypical Combinations and Scientific Impact [Article]. *Science*, *342*(6157), 468-472. https://doi.org/10.1126/science.1240474
van der Wouden, F., & Youn, H. (2023). The impact of geographical distance on learning through collaboration [Article]. *Research Policy*, *52*(2), 18, Article 104698. https://doi.org/10.1016/j.respol.2022.104698
Wagner, C. S., Whetsell, T. A., & Mukherjee, S. (2019). International research collaboration: Novelty, conventionality, and atypicality in knowledge recombination [Article]. *Research Policy*, *48*(5), 1260-1270. https://doi.org/10.1016/j.respol.2019.01.002
Wang, G. F., Gan, Y. T., & Yang, H. D. (2022). The inverted U-shaped relationship between knowledge diversity of researchers and societal impact [Article]. *Scientific Reports*, *12*(1), 10, Article 18585. https://doi.org/10.1038/s41598-022-21821-0
Wells, J. E., & Aicher, T. J. (2013). Follow the leader: A relational demography, similarity attraction, and social identity theory of leadership approach of a team's performance. *Gender Issues*, *30*(1), 1-14.
Wu, L., Wang, D., & Evans, J. A. (2019). Large teams develop and small teams disrupt science and technology. *Nature*, *566*(7744), 378-382. https://doi.org/10.1038/s41586-019-0941-9
Wu, L., Yi, F., Bu, Y., Lu, W., & Huang, Y. (2024). Toward scientific collaboration: A cost-benefit perspective. *Research Policy*, *53*(2), 104943. https://doi.org/10.1016/j.respol.2023.104943
Wuchty, S., Jones, B. F., & Uzzi, B. (2007). The Increasing Dominance of Teams in Production of Knowledge. *Science*, *316*(5827), 1036-1039. https://doi.org/10.1126/science.1136099
Xing, Y. M., Ma, Y. F., Fan, Y., Sinatra, R., & Zeng, A. (2025). Academic mentees thrive in big groups, but survive in small groups [Article; Early Access]. *Nature Human Behaviour*, 18. https://doi.org/10.1038/s41562-025-02114-8

Xu, F. L., Wu, L. F., & Evans, J. (2022). Flat teams drive scientific innovation [Article]. *Proceedings of the National Academy of Sciences of the United States of America*, *119*(23), 3, Article e2200927119. https://doi.org/10.1073/pnas.2200927119

Xu, H. M., Bu, Y., Liu, M. J., Zhang, C. W., Sun, M. Y., Zhang, Y., Meyer, E., Salas, E., & Ding, Y. (2022). Team power dynamics and team impact: New perspectives on scientific collaboration using career age as a proxy for team power [Article]. *Journal of the Association for Information Science and Technology*, *73*(10), 1489-1505. https://doi.org/10.1002/asi.24653

Xu, H. M., Liu, M. J., Bu, Y., Sun, S. J., Zhang, Y., Zhang, C. W., Acuna, D. E., Gray, S., Meyer, E., & Ding, Y. (2024). The impact of heterogeneous shared leadership in scientific teams [Article]. *Information Processing & Management*, *61*(1), 13, Article 103542. https://doi.org/10.1016/j.ipm.2023.103542

Yang, A. J., Freeman, R. B., & Deng, S. (2026). A semantic atlas of journals: Structure, position, and dispersion. *Journal of the Association for Information Science and Technology*.

Yang, X., Zhou, X., & Cao, C. (2024). Beamtimes and knowledge production times: how big-science research infrastructures shape nations' domestic and international science production. *Humanities and Social Sciences Communications*, *11*(1), 1462. https://doi.org/10.1057/s41599-024-03993-4

Yang, Y., Tian, T. Y., Woodruff, T. K., Jones, B. F., & Uzzi, B. (2022). Gender-diverse teams produce more novel and higher-impact scientific ideas [Article]. *Proceedings of the National Academy of Sciences of the United States of America*, *119*(36), 8, Article e2200841119. https://doi.org/10.1073/pnas.2200841119

Yoo, H. S., Jung, Y. L., Lee, J. Y., & Lee, C. (2024). The interaction of inter-organizational diversity and team size, and the scientific impact of papers [Article]. *Information Processing & Management*, *61*(6), 15, Article 103851. https://doi.org/10.1016/j.ipm.2024.103851

Zeng, A., Fan, Y., Di, Z. G., Wang, Y. G., & Havlin, S. (2022). Impactful scientists have higher tendency to involve collaborators in new topics [Article]. *Proceedings of the National Academy of Sciences of the United States of America*, *119*(33), 9, Article e2207436119. https://doi.org/10.1073/pnas.2207436119

Zhang, L., Cao, Z., Shang, Y. Y., Sivertsen, G., & Huang, Y. (2024). Missing institutions in OpenAlex: possible reasons, implications, and solutions [Article; Early Access]. *Scientometrics*, *129*(10), 5869–5891. https://doi.org/10.1007/s11192-023-04923-y

ZHANG, M., FAN, X., LI, Y., & LI, Z. (2024). Facilities' scientists affect scientific impacts of large-scale research infrastructures' outputs. *Bulletin of Chinese Academy of Sciences (Chinese Version)*, *40*(8), 1357-1369.

Zhang, M., Li, Y., Li, Y., & Li, Z. (2026). Scientific tools and Innovation: Big Science Facilities Yield More Novel and Interdisciplinary Knowledge. *arXiv preprint arXiv:2604.19396*.

Zhang, M., Lyu, P., Li, Y., & Li, Z. (2025). *Scientific Travelers Associated with Less Disruption but Better Scientific Novelty* 20th International Conference on Scientometrics and Informetrics (ISSI 2025), Yerevan, Armenia. https://doi.org/10.51408/issi2025_085

Zhang, M., Wang, L., & Li, Z. (2025). The impact of team compositions on disruptive and novel research in large-scale research infrastructures. *Scientometrics*, *130*(5), 2987–3011. https://doi.org/10.1007/s11192-025-05319-w

Zhang, M., Wang, L., Zhang, L., & Li, Z. (2026). Co-utilizing global big science facilities: a novel type collaboration and the impacts on scientific disruption. *Humanities and Social Sciences Communications*. https://doi.org/10.1057/s41599-026-06992-9

Zhang, M.-Z., Wang, T.-R., Lyu, P.-H., Chen, Q.-M., Li, Z.-X., & Ngai, E. W. T. (2024). Impact of gender composition of academic teams on disruptive output. *Journal of Informetrics*, *18*(2), 101520. https://doi.org/10.1016/j.joi.2024.101520